\documentclass[letterpaper, 10 pt, conference]{ieeeconf}  % Comment this line out if you need a4paper

\IEEEoverridecommandlockouts                              % This command is only needed if 
\usepackage{graphics} % for pdf, bitmapped graphics files
\usepackage{caption}
\usepackage{float} 
\usepackage{subcaption}
\usepackage{epsfig} % for postscript graphics files
\usepackage{mathptmx} % assumes new font selection scheme installed
\usepackage{times} % assumes new font selection scheme installed
\usepackage{amsmath} % assumes amsmath package installed
\usepackage{amssymb}  % assumes amsmath package installed
\usepackage{bm}

\usepackage[mathscr]{eucal}
\usepackage{algpseudocode}
\usepackage{algorithmicx}
\usepackage{algorithm}
\usepackage{booktabs}

\graphicspath{{./figs/}}
\DeclareGraphicsExtensions{.pdf,.jpeg,.png,.jpg}

\renewcommand{\algorithmicrequire}{\textbf{Input:}}
\renewcommand{\algorithmicensure}{\textbf{Output:}}

\title{An Extendable Maneuver Management Framework with Fault-Tolerant Mechanism for Vehicle Platoon Control System in Highway Scenario}

\author{Chang Liu$^{1}$, Yugong Luo$^{1}$, Pengfei Li$^{1}$, Chunhui Xing$^{2}$, and Weiwei Kong$^{*2}$% <-this % stops a space
\thanks{*This study is supported by the National Natural Science Foundation of China under Grant 52002209 and Grant 51975310.}
\thanks{$^{1}$Chang Liu, Yugong Luo and Pengfei Li are with the School of Vehicle and Mobility, Tsinghua University, Haidian, Beijing 100084, P.R China.
}
\thanks{$^{2}$ Chunhui Xing and Weiwei Kong are with the College of Engineering, China Agricultural University, Haidian, Beijing 100083, P.R China. Weiwei Kong is the corresponding author,
{\tt\small kongweiwei@cau.edu.cn} }
}

\begin{document}

\maketitle
\thispagestyle{empty}
\pagestyle{empty}

%%%%%%%%%%%%%%%%%%%%%%%%%%%%%%%%%%%%%%%%%%%%%%%%%%%%%%%%%%%%%%%%%%%%%%%%%%%%%%%%
\begin{abstract}
Vehicle platoon often face the problem of lack of scalability of maneuvers in practical applications. Once a new scenario is added, the original program may no longer be available.
To deal with this problem, this paper introduces a two-dimensional maneuver management framework with fault-tolerant mechanism on the basis of the proposed hierarchical architecture for the platoon control system. Maneuvers and roles are two dimensions, based on which the management strategies are decoupled. This makes each vehicle in the platoon has the ability to execute management strategies of various maneuvers and the new maneuver could be extended without revising the existing part. 
The fault-tolerant mechanism is designed as a maneuver triggered by hardware failures to keep safe before taking over. 
Furthermore, three typical maneuvers are selected for case studies to illustrate how the management strategies in this framework work. 
Finally, a comprehensive simulation scenario integrating different maneuvers is designed and a real-world implementation using micro-vehicles is conducted. Result shows that the proposed two-dimensional framework could effectively deal with various maneuvers and satisfy the computational real-time requirements.

\end{abstract}

%%%%%%%%%%%%%%%%%%%%%%%%%%%%%%%%%%%%%%%%%%%%%%%%%%%%%%%%%%%%%%%%%%%%%%%%%%%%%%%%
\section{INTRODUCTION}
\label{sec:introduction}
Intelligent Transport Systems(ITS) will play an important role in the future of transportation, such as estimating the speed of road vehicles accurately \cite{fernandez_llorca_vision-based_2021}. Vehicle platoon is also a promising application for ITS, which has been proposed with the objectives of enhancing safety\cite{xu_coordinated_2013}, improving highway utility \cite{dao_strategy_2013}, and increasing fuel economy\cite{li_stabilizing_2017}. The first research project of the platoon was first demonstrated by PATH in the 1980s\cite{shladover_automated_1991}, and further developed by variable research projects, namely Chauffeur\cite{tsugawa_review_2016}, KONVOI\cite{kunze_organization_2011}, EnergyITS\cite{tsugawa_results_2014}, and TROOP\cite{lee_novel_2020}. These projects verified the feasibility of platoon and the ability to deal with specific maneuvers, such as joining the platoon or steady platooning. However, when considering practical applications, the platoon needs to switch between different maneuvers including some emergency maneuvers. The new maneuvers should be added to the current platoon control system without revising the existing strategies. As a result, it is important to design an extendable maneuver management framework to cope with the various maneuvers in the actual scenario.

Significant energy savings are achieved for the distance below $20m$ between vehicles in the platoon, which means a headway time of $0.8s$ at $25m/s$ \cite{bijlsma_fail-operational_2017}. Short inter-vehicle distance is the key to traffic efficiency and fuel consumption for the platoon, but this makes the safety requirements challenging. Such short following distance makes it hard for the driver to take over if a hardware failure occurs. The current V2V communication system is not very reliable, and there will be failures such as unbearable time delay and disconnection. Meanwhile, the millimeter-ware radar may also fails during the platoon operates, then the sensing distance may have a sudden change. When any of the hardware fails, it is not enough to simply issue a takeover request to the driver, which may cause chain collisions. Therefore, a fault-tolerant safety mechanism for the platoon control system is necessary. 

To handle the problems the platoon may meet when driving in the actual scenario, it is important to design a novel maneuver management framework with fault-tolerant mechanism, which is capable of dealing with various maneuvers in practical applications and maintaining the safety of the platoon in the presence of the hardware failures. The framework is extendable for adding new maneuvers, and also universal for each role in the platoon. Besides, the fault-tolerant mechanism is required to switch to the suitable controller that could remain acceptably functions after sending the takeover request to the driver when the hardware fails. Transitioning into the controller that could remain acceptably functions in the presence of failures is referred in this paper as \textit{functionality degradation}.

The main contributions are as follows:
\begin{enumerate}
        \item The proposed extendable framework decouples the management strategies for different maneuvers, which brings the benefits that each maneuver is independent of the others, so the maneuvers could be extended without revising the existing part. 
        \item The fault-tolerant mechanism is applied in the management framework. The current controller will be switched into the degrading controller to stay safe until the driver takes over when the hardware failure occurs.
\end{enumerate}

The remainder of this paper is structured as follows: Section \uppercase\expandafter{\romannumeral2} introduces the related work of this paper. Section \uppercase\expandafter{\romannumeral3} proposes the hierarchical architecture for the platoon control system. Section \uppercase\expandafter{\romannumeral4} introduces the two-dimensional maneuver management framework and explains the detailed illustrations of maneuver management strategies. Section \uppercase\expandafter{\romannumeral5} illustrates the integrated simulation result of different maneuvers. Last, the results of the field implementation is presented in \uppercase\expandafter{\romannumeral6}.

\section{RELATED WORK}
The researches concerning the vehicle platoon have become very popular in recent years. The methods of platoon stability control have been widely studied and well developed \cite{gao_robust_2016}. However, there are problems that need to be solved if the automated platoon could drive in the real-world traffic. 

Some researches concentrate on the maneuver management strategy, which is mainly about how different maneuvers are switched and how the specific maneuvers are implemented. The other researches focus on the fault-tolerant mechanism that is the key to deal with the hardware failures in the platoon. The following parts will introduce the related work of these two areas.

\subsection{Maneuver Management Strategy}

In the researches related to vehicle platoon implementing management strategies, the hierarchical architecture is widely adopted. \cite{horowitz_control_2000} summarized The multilayer automated highway system (AHS) control architecture, including network, link, coordination, regulation, and physical layer from top to bottom is summarized. \cite{kazerooni_interaction_2015} presented their vehicle control system using three layers, namely perception, control, and supervisory layer, based on the proposed interaction protocols. \cite{ploeg_cooperative_2018} adopted a four-layered control architecture in which the tactical layer hosts the interaction protocols accommodating common highway maneuvers and urban applications. \cite{huang_path_2019} proposed a hybrid system consisting of three parts: information processing, hybrid automata for cooperative maneuver switching, and vehicle control.

To simplify the description of different maneuver management strategies, some papers have defined the platoon maneuvers and management strategies formally. Three basic platooning maneuvers including merge, split, and lane change, and a set of micro-commands to accomplish these maneuvers were summarized in \cite{amoozadeh_platoon_2015}. An agent-based method where each of the maneuvers involves a different set of controller agents to decompose the complex tasks was employed in \cite{kazerooni_interaction_2015}. An ontological model of platooning objects, platoon properties and abstract basic building blocks of platoon operations was proposed in \cite{maiti_conceptualization_2017}. And a hierarchical state machine that simplifies the maneuver design process was designed in \cite {ivanchev_hierarchical_2021}. These studies decomposed the maneuvers of the platoon, which makes implement the maneuver easier. However, none of them explicitly suggest the idea of extendable maneuver management strategies. 

In addition to the maneuver management strategies, researchers have proposed specific cases of cooperative maneuvers. \cite{badnava_platoon_2021} summarized the recent studies on platoon maneuvers and most of them focused on join and lane change maneuver. In \cite{amoozadeh_platoon_2015},\cite{ivanchev_hierarchical_2021}, \cite{zhang_v-pada_2011}, the strategy was implemented through finite state machine describe the operating process of different maneuvers. Some researches focused on the specific platoon maneuver and optimized it for traffic efficiency and fuel consumption. \cite{maiti_impact_2020} introduced three different merge vehicle selection methods including front, middle, and tail merge and analyzed the impact on average traffic speed with varied speed adjustment strategy and the traffic density. They also studied two types of ad-hoc platoon formation and platoon dissolution strategies in \cite{maiti_ad-hoc_2021}. And the platoon lane change strategies can be seen in \cite{nie_cooperative_2022}. In \cite{fida_improved_2021}, the management strategy for multiple vehicle joining and leaving the platoon simultaneously was proposed. Some studies have conducted the strategy validation. \cite{braud_avdm_2021} implemented the model in the BEHAVE mixed traffic simulation tool. And \cite{calvert_cooperative_2020} implemented a field operational test of the platoon in real traffic.

\subsection{Fault-Tolerant Mechanism}

The V2V failure and the radar failure is the hazardous event in the platoon system. \cite{stolte_taxonomy_2021} distinguished between different fault-tolerant mechanism. It's a fail-degraded system if it can provide its specified functionality with below nominal performance. And it's a fail-operational system if it can provide its specified functionality with at least nominal performance. 

Some researches focus on the the fail-degraded mechanism. When a hazard occurs, the system will switch to a degraded controller and cannot maintain the original control performance. \cite{hasan_fault-tolerant_2019} monitored the communication status in a platoon and switches between three typical controllers. \cite{sljivo_assuring_2017} proposed the safety assurance of degradation cascades using contracts, which could cope with not only the local failures, but also ones in other vehicles in the platoon by V2V communication. \cite{harfouch_adaptive_2018} designed an adaptive switched control strategy, which selects the augmented CACC controller or the augmented ACC controller depending on the quality of communication.

Except for fail-degraded mechanism, some researches give the fail-operational mechanism which maintain the nominal performance using some extra methods. \cite{van2016sensor} discussed the V2V and sensor failures and give the backup sources if the desired information cannot be obtained. \cite{gratzer_string_2022} proposed an extended time gap spacing policy, which provides safe platoon operation with robust string stability without V2V communication.

The first part of the researches have disscussed how management strategies are formalized and how the specific maneuver is implemented, none of them consider the scalability of maneuvers in practical applications. 
The second part of the researches focus on the degradation and recovery of system functionality depending on the degree of the hardware failure. They also discuss the method to maintain the system performance by designing fault-tolerant controllers. However, they have not pointed out that the driver takeover is the last safety guarantee if the failure lasts long enough. The radar and V2V failures are assumed to be permanent in this paper, so the recovery of functionality degradation is not considered.

In this paper, the scalability of maneuvers and the response to hardware failures are considered at the same time and then a novel maneuver management framework with fault-tolerant mechanism is proposed, which is able to deal with the maneuvers not included in the existing strategies, and maintain the safety of the platoon when the hardware fails.

\section{HIERARCHICAL ARCHITECTURE FOR VEHICLE PLATOON CONTROL SYSTEM}

\begin{figure}[thpb]
        \centering
        %\framebox{\parbox{3in}{hhh}}
        \includegraphics[scale=0.4]{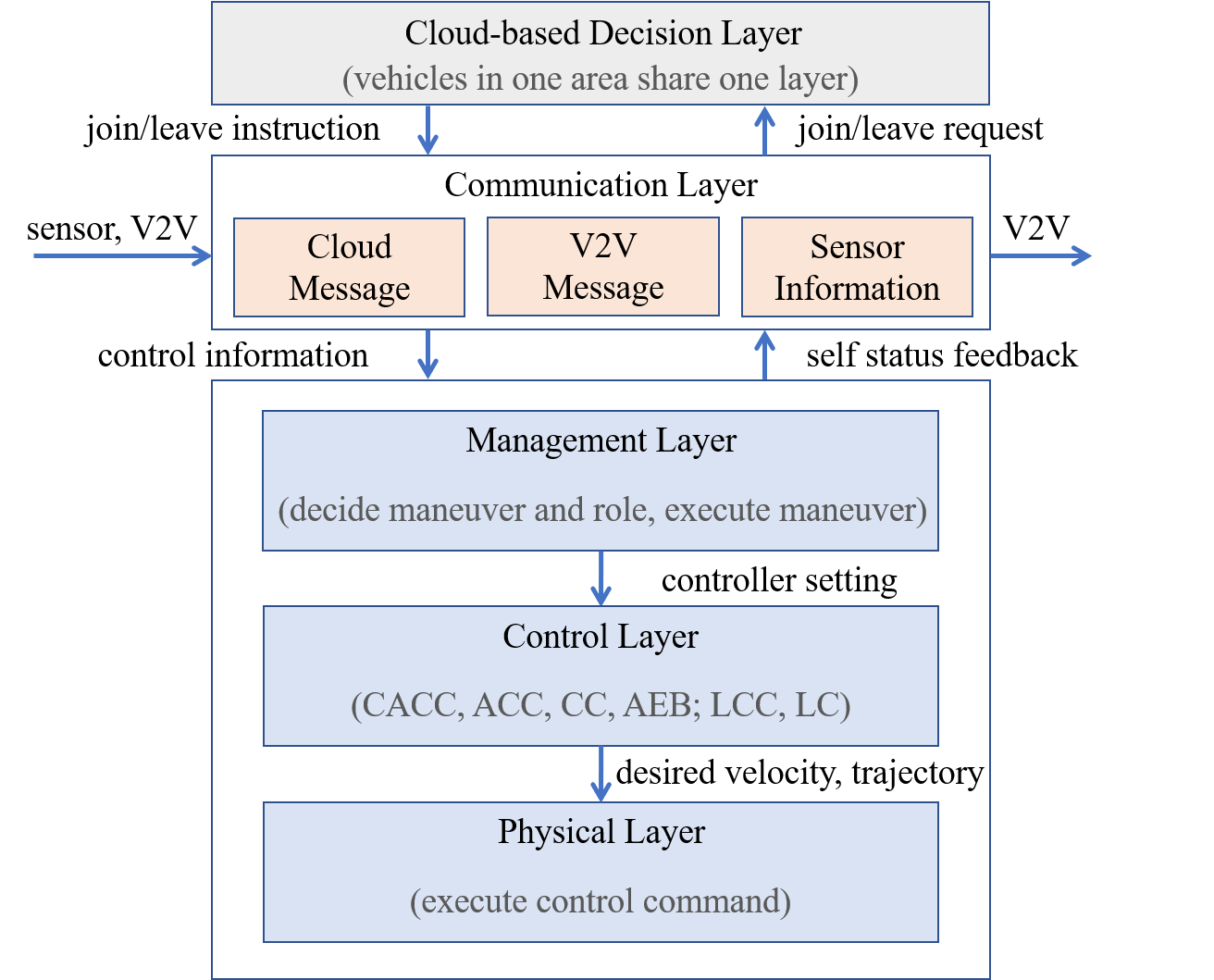}
        \caption{Hierarchical architecture for platoon control system.}
        \label{overview}
     \end{figure}

A hierarchical architecture for the vehicle platoon control system is presented as shown in Fig.~\ref{overview}, consisting of five layers ranging from the \textit{Cloud-Based Decision Layer} to the \textit{Physical Layer}. In this section, the function of the five layers will be explained in detail.

\subsubsection{Cloud-Based Decision Layer} Supposing the platoon control system is supported by the cloud platform, this layer could perceive the current traffic information around the specific area through the sensors on the participating vehicles and road-side infrastructure. Besides, this layer can also obtain the travel schedules and receive the join or leave request from vehicles which is responsible for optimizing the travel time and traffic flow. It determines whether and when a vehicle should join or leave. After that, it sends the control information leveraging on wireless communication. The cloud-based decision layer will be implemented as a centralized control center, so the vehicles in one area share this layer and control information. This layer will be referred to as the cloud layer in the following.

\subsubsection{Communication Layer} This layer processes the information from three different sources, which are cloud message receiving from the cloud layer, V2V message from other autonomous vehicles, and sensor information from the sensors aboard (which are radars and cameras in this paper). This layer is implemented on each vehicle and sends or receives the necessary messages within the range of V2V communication. 

\subsubsection{Management Layer} This layer, which is the focus of this paper, contains the logic of deciding which maneuver to take and selecting the platoon role using the current control information. After that, the management layer will guide the vehicle to perform the maneuver, which will be explained in the next section in detail. In the process of the maneuver, this layer also determines the controller it needs to use and sends the controller setting information to the control layer. Besides, the fault-tolerant mechanism is considered in this layer and the corresponding management strategy will be performed when the hardware failures are detected. 

\subsubsection{Control Layer} This layer is responsible for calculating the desired velocity and trajectory using the longitudinal and lateral controllers selected by the management layer, where the longitudinal controllers include Cooperative Adaptive Cruise Control (CACC), Adaptive Cruise Control (ACC), Cruise Control (CC) and Autonomous Emergency Braking (AEB), while the lateral controllers consist of Lane Center Control (LCC) and Lane Change (LC). This layer contains not only the CACC controller that is used for steady platooning, but also some simpler controllers, such as ACC and CC. When the hardware failures happen, the system could switch to the degraded controller for temporary safety before the driver to take over. Because the control algorithms of the controllers mentioned are mature, the implementation of these controllers will not be discussed in detail. 

\subsubsection{Physical Layer} This layer executes the control command including desired velocity and local trajectory from the upper layer and operates the vehicle directly. Different from the communication layer, the management layer, and the control layer which are exactly the same in every vehicle, the physical layer varies between different brands of participating vehicles due to different dynamics.

For distinct maneuvers, different management strategies need to be designed in the management layer, but the other four layers are not directly correlated with the maneuvers. Therefore, this paper focuses on the \textit{Management Layer}.

\section{MANAGEMENT LAYER WITH TWO-DIMENSIONAL FRAMEWORK} 

There are a variety of maneuvers to deal with in real-world application. In one maneuver, different roles of platoon vehicles will execute the different management strategies. All of strategies will be implemented with the maneuver management framework in the \textit{Management Layer}. In this section, a detailed explanation of the two-dimensional maneuver management framework will be first given. Then this section will introduce the logic of maneuver deciding, and role selecting. After that, Join Middle and AEB Head are taken as example to illustrate the management strategy under the specific maneuver and role. Finally, the fault-tolerant mechanism is explained. 

\begin{figure}[thpb]
        \centering
        %\framebox{\parbox{3in}{hhh}}
        \includegraphics[scale=0.37]{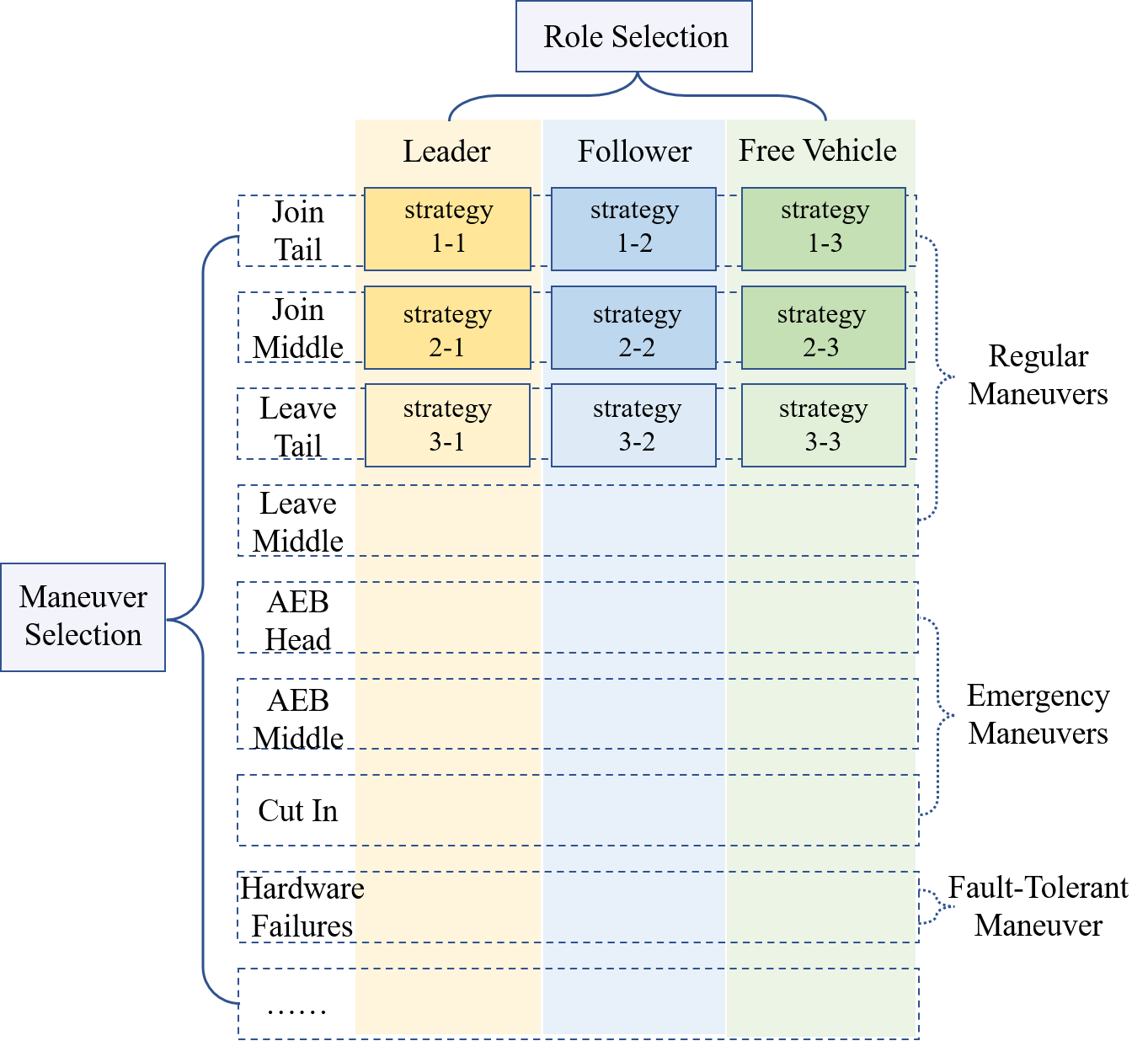}
        \caption{The extendable two-dimensional framework for the management layer.} 
        \label{management}
\end{figure}

Fig.~\ref{management} shows an overview of the two-dimensional framework for the management layer. Each row represents a maneuver, and each column represents a role including the free vehicle, the leader, and the follower. The leader is the car driving at the front of the entire platoon and it manages the information of the platoon such as platoon size and ID series. The follower is the car that is already a member of the platoon following the velocity of the leader. And the free vehicle is the car that is controlled by the driver temporarily, and it will be controlled by the control layer once it joins the platoon.
By using the control information, this layer decides the maneuver to take and selects a role among the three. Only one maneuver and only one role can be selected by each vehicle at a time. Subsequently, the vehicle performs the management strategy under the specific maneuver and role.

The existed maneuvers listed in Fig.~\ref{management} could be divided into three categories. The first is the regular maneuvers usually encountered in platoon travel, including Join Tail, Join Middle, Leave Tail, and Leave Middle. The second is the emergency maneuvers caused by the intruding object that cuts in suddenly, including AEB Head, AEB Middle, and Cut In. The last is the fault-tolerant maneuver which is triggered when the hardware fails. The last row of the figure represents the extendable maneuvers that could be added into the framework if needed.

Each strategy generally includes these actions: sending or receiving messages, updating platoon information, waiting for certain events, choosing appropriate controller, and transitioning the role in the platoon. The concrete processes of the management strategies would be illustrated in the following subsections. 

Compared with the researches mentioned in the previous section, this framework has two advantages: extendable and universal.

\begin{itemize}
        \item Extendable: To deal with the actual driving scenario, the maneuvers under this framework could be extended, rather than limited to several given maneuver strategies.  
        \item Universal: The management framework equipped in every participating vehicle is exactly the same whether it is the leader, the follower, or the free vehicle, so the framework is universal for all the roles in the platoon.
\end{itemize}

\subsection{Maneuver Selection}

\begin{figure}[thpb]
        \centering
        %\framebox{\parbox{3in}{hhh}}
        \includegraphics[scale=0.27]{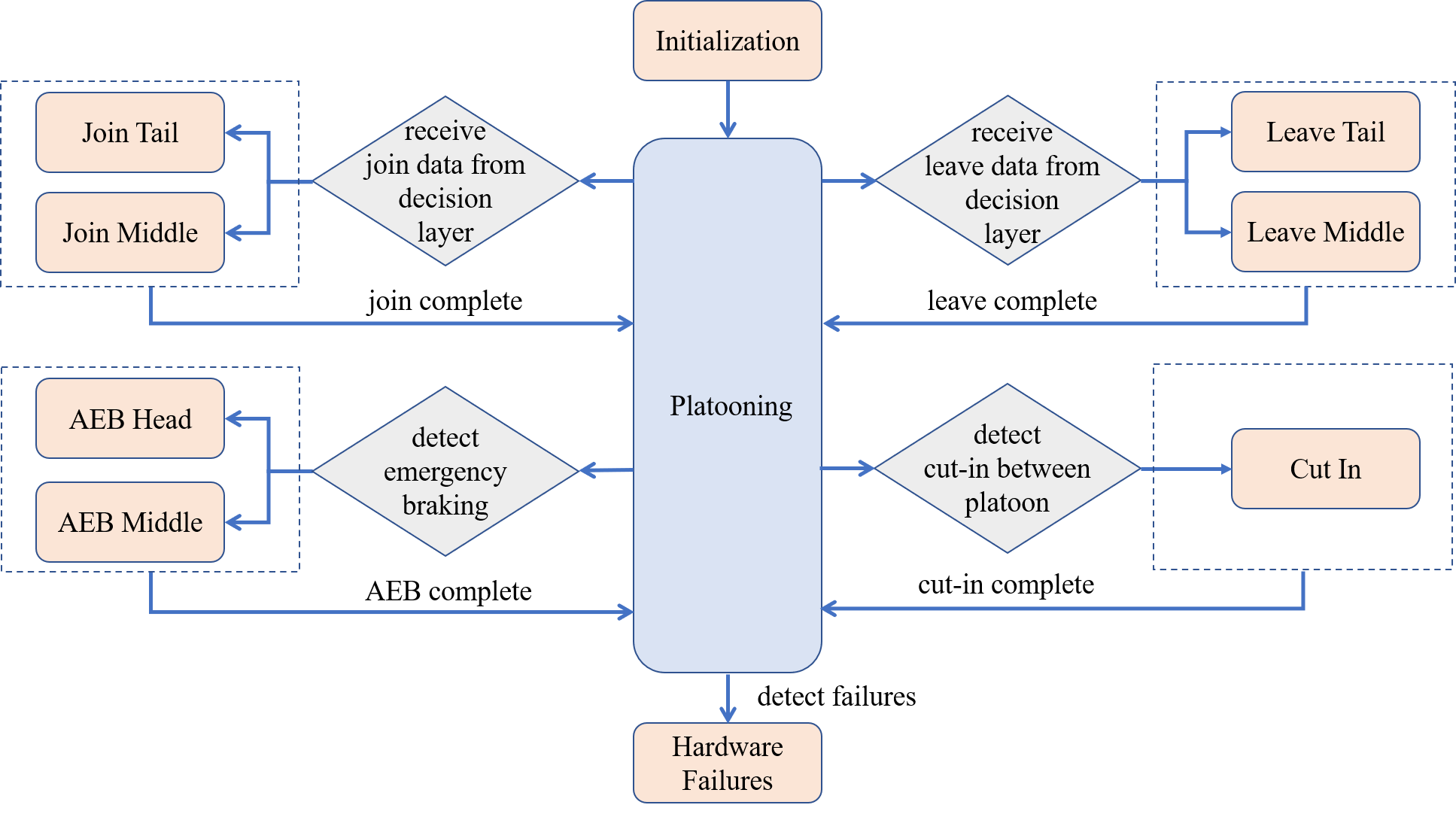}
        \caption{Finite State Machine (FSM) of maneuver selection. After the maneuver is completed, the state of the vehicle returns to the platooning state.} 
        \label{maneuver}
     \end{figure}

The logic of maneuver selection could be expressed as the Finite State Machine(FSM) as shown in Fig.~\ref{maneuver}. In this paper, it is supposed that the vehicle keeps in platooning state after initializing unless the following three kinds of events happen and then the state will transition. 

\subsubsection{Receiving instruction from the cloud layer} Taking the {Join} maneuver as an example, the cloud layer analyzes the global information and calculates the appropriate time and location for the free vehicle to join in the existing platoon. Then the cloud layer transmits the join instruction to related vehicles. After receiving the instruction, the vehicle transitions to the {Join} maneuver. It is classified into {Join Tail} and {Join Middle} maneuver depending on whether Join happens at the end of the platoon. After the Join maneuver is completed, the vehicle returns to the platooning state again. The same process will happen with {Leave} maneuver.

\subsubsection{Detecting obstacles from sensors} Once the sensors aboard detect the sudden appearances of obstacle vehicles on the lane satisfying the Time to Collision (TTC) condition, the ego vehicle performs emergency braking, causing all the vehicles behind it to undergo emergency braking. It is supposed that the vehicle enters the {AEB} maneuver in this paper. It is classified into {AEB Head} and {AEB Middle} depending on the location of obstacles. If the sensors detect a vehicle cuts in the existing platoon without negotiating while not satisfying the TTC condition, it is supposed that the ego vehicle enters {Cut In} maneuver. In this case, the behavior of the cut-in vehicle is uncertain for the platoon. As a result, the controllers of the vehicles behind the cut-in vehicle need to change to increase the gap to assure safety.

\subsubsection{Detecting hardware failures} There is a module performing fault detection in the control system. If the module detects the hardware failure, it will broadcast this information. The module could also monitor the communication ability of other vehicles. Because the hardware failures are assumed to be permanent in this paper, the recovery is not considered. After the functionality degradation, the driver will take over the vehicle and this vehicle is no longer a member of the platoon.

There is one thing to note that once a vehicle in the platoon transitions to one maneuver state, then the other vehicles will also select this maneuver. To achieve this, the vehicle already chooses the maneuver will broadcast its selection and the vehicles that are still in the platooning state will select this maneuver. However, if the V2V device on the car fails, it will lose its ability to broadcast. As a result, the other vehicles need to detect this faulty vehicle by themselves and then transition to then Hardware Failures maneuver. 

\subsection{Role Selection}
\begin{figure}[thpb]
        \centering
        %\framebox{\parbox{3in}{hhh}}
        \includegraphics[scale=0.45]{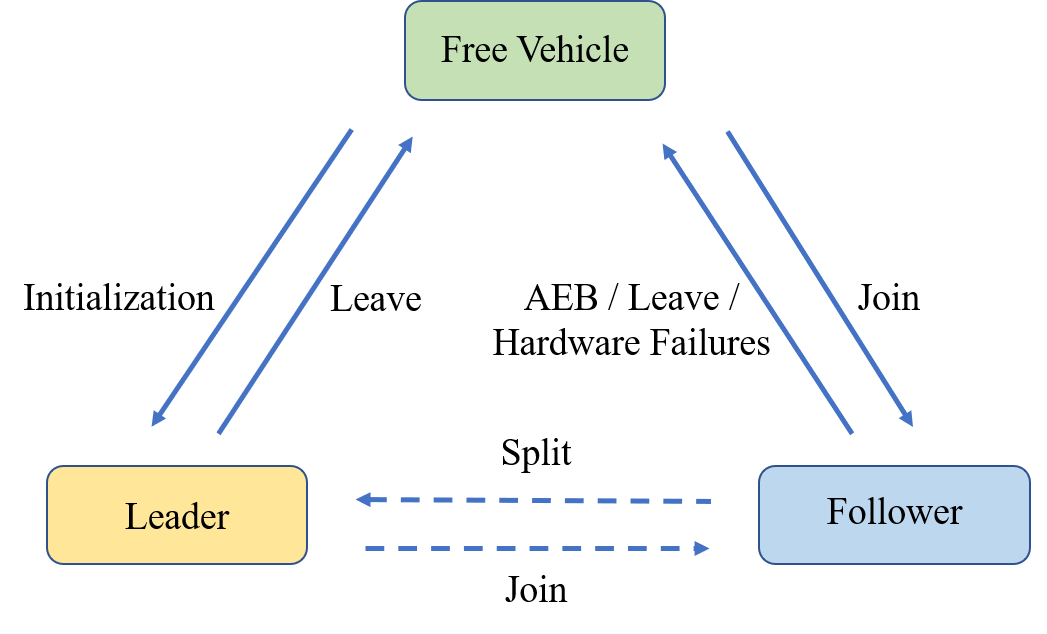}
        \caption{Finite State Machine (FSM) of role selection. The dashed lines of transition, which are not explained in this paper, are shown here merely for completeness.} 
        \label{role}
     \end{figure}

The logic of role selection could be expressed as another FSM as shown in Fig.~\ref{role}. The role of a vehicle has three possible options: the free vehicle, the leader, and the follower. Different roles perform different functions in a maneuver. As shown in Fig.~\ref{role}, each role could be transformed into any other one through a specific maneuver. After receiving the instruction from the management strategy to change the role (usually happens at the end of the management strategy process), the state of the role will transition. 

All the participating vehicles are free vehicle at the beginning, and after the initialization one of them could transition to a leader depending on the system settings. A leader remains its role until it finally leaves the platoon or joins another platoon from behind which is not concerned in this paper. A free vehicle could become a follower through the Join maneuver while a follower could return to a free vehicle through AEB, Leave or Hardware Failures maneuver. Besides, a follower could also become a leader during Split maneuver which is also not included in this paper.

\subsection{Maneuver Management Strategy Design}
The management strategies of the typical maneuvers shown in Fig.~\ref{management} are designed. This section will take the Join Tail maneuver, AEB Head maneuver and Fault-tolerant maneuver as three examples to illustrate the details of management strategies.

\subsubsection{Join Tail Maneuver}
\renewcommand{\algorithmicrequire}{\textbf{Input:}}  % Use Input in the format of Algorithm
\renewcommand{\algorithmicensure}{\textbf{Output:}} % Use Output in the format of Algorithm

Algorithm~\ref{JTFREE}, and~\ref{JTLEAD} illustrate the strategy of {Join Middle}  maneuver when a free vehicle receives the order from the cloud layer and joins the existing platoon in the middle. The strategies for different roles are connected and synchronized through V2V communication. 

\begin{algorithm}[thpb]
        \caption{Join Tail for the Free Vehicle}
        \label{JTFREE}
        \begin{algorithmic}[1]
        \Require
        1. UpdateFlag: message received from the leader. 2. instruction from the cloud layer. 3. $D$: distance from ego vehicle to the front vehicle in the same lane.
        \Ensure
        1. JoinFlag: message sent to the leader. 2. $C_{1}$: Longitudinal Controller. 3. $R$: selected role  
        \State JoinFlag = 0;
        \If {receive Join instruction}
        \State $C_{1}$ = ACC; 
        \While {$D > 30$}
        \State Wait;
        \EndWhile
        \State JoinFlag = 1;
        \While {UpdateFlag == 0}
        \State Wait;
        \EndWhile
        \State $C_{1}$ = CACC;
        \State $R$ = Follower;
        \EndIf
        \end{algorithmic}
\end{algorithm}

\begin{algorithm}[h]
        \caption{Join Middle for the Leader} 
        \label{JTLEAD}
        \begin{algorithmic}[1]
    \Require
    1. JoinFlag: message received from joining vehicle.
    \Ensure
    1. UpdateFlag: message sent to the free vehicle. 2. updated size and ID series  
    \State UpdateFlag = 0;   
    \While {JoinFlag == 0}
    \State Wait;
    \EndWhile
    \State Update size;
    \State Update ID series;
    \State UpdateFlag = 1;
        \end{algorithmic}
\end{algorithm}

The normal driving speed of the platoon is set to $20 m/s$. After receiving the join instruction, the follower which is exactly behind the join location will set the speed to $15m/s$ for increasing the distance to the front vehicle. When the distance reaches $30m$, it will send \textit{EvadeFlag} to the free vehicle and set the speed back to $20m/s$. Then the free vehicle will change its lane and join the platoon, after which it will send \textit{JoinFlag} to other vehicles. Finally, the leader will update the information of platoon members.

\subsubsection{AEB Head Maneuver}
Algorithm~\ref{AEBFOLLOW} and Algorithm~\ref{AEBLEAD} illustrate the strategy of {AEB Head} when an intruding vehicle changes its lane suddenly before the leading vehicle and the platoon vehicles perform emergency braking. Because it is supposed that the free vehicle needs no extra action, this part only contains the strategies for the leader and the free vehicle. When this maneuver ends, the cloud layer will send join instruction and the platoon will execute the Join maneuver subsequently.

\begin{algorithm}[h]
        \caption{AEB Head for the Follower}
        \label{AEBFOLLOW}
        \begin{algorithmic}[1]
          \Require
          1. SafeFlag: message received from the leader. 2. UpdateFlag: message received from the leader 3. v: velocity of ego vehicle
          \Ensure
          1. join request sent to the cloud layer. 2. $C_{1}$: Longitudinal Controller. 3. $R$: selected role
          \While{v $\ge$ 0}
          \State $C_{1}$ = AEB;
          \EndWhile
          \State $C_{1}$ = driver;
          \While {SafeFlag == 0}
          \State Wait;
          \EndWhile  
          \State Send Join Request;   
          \While {UpdateFlag == 0}
          \State Wait;
          \EndWhile
          \State $R$ = FreeVehicle;
        \end{algorithmic}
\end{algorithm}
      
\begin{algorithm}[h]
              \caption{AEB Head for the Leader}
              \label{AEBLEAD}
              \begin{algorithmic}[1]
                \Require
              1. sensor data.
          \Ensure
          1. SafeFlag: message sent to the free vehicle 2. UpdateFlag: message sent to the free vehicle. 3. updated size and ID series 4. $C_{1}$: Longitudinal Controller.
          \State UpdateFlag = 0;
          \State SafeFlag = 0;
          \State $C_{1}$ = AEB;   
          \While {obstacle remains}
          \State Wait;
          \EndWhile
          \State SafeFlag = 1;
          \State Update size;
          \State Update ID series;
          \State UpdateFlag = 1;
              \end{algorithmic}
\end{algorithm}      

When switching to the AEB maneuver, the leader adopts the AEB controller unless the intruding car leaves. Then it sends \textit{SafeFlag} to the followers and updates the platoon information. At the same time, the followers also adopt the AEB controller. When receiving \textit{SafeFlag}, they are restarted by drivers and send the Join request to the cloud layer.

\subsubsection{Fault-Tolerant Mechanism Design}
The V2V failure and the radar failure is the hazardous event in the platoon system. When the hardware failure occurs, the platoon control system needs to react immediately due to the close spacing of the vehicles. If only issue a takeover request to the driver while the control system does not react, it may cause chain collisions because the reaction time of driver is not fast enough. 

Therefore, a fault-tolerant mechanism is required to switch to another suitable platoon controller that no longer uses the information from the fault hardware. By switching between existing controllers and adjusting inter-vehicle distances, the control system maintains the temporary safety of the platoon to wait for the driver to takeover. Functionality degradation is considered as the way to maintain the safety of the platoon in the presence of failures.

The Hardware Failures maneuver could be triggered when detecting the failures happened on itself, monitoring the probblem of the V2V device on other vehicles, or receiving the fault broadcast from others. Algorithm~\ref{HDFOLLOW} and Algorithm~\ref{HDLEAD} illustrate the strategy of {Hardware Failures}. Because it is assumed that hardware failures only occur while steady platooning, the strategy of the free vehicle is not involved in this part.

\begin{algorithm}[h]
        \caption{Hardware Failures for the Follower}
        \label{HDFOLLOW}
        \begin{algorithmic}[1]
          \Require
          1. FaultFlag: message received from other platoon vehicles. 2. sensor and V2V data.
          \Ensure
          1. takeover request. 2. $C_{1}$: Longitudinal Controller. 3. $R$: selected role
          \If{radar fails}
          \State Send takeover request;
          \State $C_{1}$ = CC;
          \EndIf
          \If{V2V communication fails and radar works}
                \State Send takeover request;
                \State $C_{1}$ = ACC;
          \EndIf
          \If{FaultFlag == 1 or the faulty vehicle detected}
                \If{behind the faulty vehicle}
                \State Send takeover request;
                \State $C_{1}$ = ACC;
                \EndIf
          \EndIf
          \If{$C_{1}$ == ACC or $C_{1} == CC$}
                \While{not taken over}
                \State Wait;
                \EndWhile
                \State $R$ = FreeVehicle;
          \EndIf
        \end{algorithmic}
        \break
\end{algorithm}

\begin{algorithm}[h]
        \caption{Hardware Failures for the Leader}
        \label{HDLEAD}
        \begin{algorithmic}[1]
        \Require
        1. FaultFlag: message received from other platoon vehicles.
        \Ensure
        1. updated size and ID series
        \State Update size;
        \State Update ID series;
        \end{algorithmic}
\end{algorithm} 

If the follower switches into this maneuver, it first check if the radar fails. If so, it sends takeover request and chooses the CC controller for lack of the distance information. Then if the V2V device fails, it chooses the ACC controller because V2V communication is not reliable. 
When the radar or the V2V device both work, there is the failure happened on other vehicles. If the vehicle is behind the faulty vehicle, then it also chooses ACC controller to increase the inter-vehicle distance and wait for the driver to takeover. When this maneuver ends, the faulty vehicle and the vehicles following the faulty vehicle are all taken over by the drivers. Because the leader is operated by the driver, there is no need for it to perform functionality degradation.

\section{SIMULATION RESULTS}

In this section, we first introduce the settings of the simulation. Then we show the overall platoon speed trajectories of the integrated maneuvers and the analyze the specified results of platoon performance. Last we validate the fault-tolerant mechanism in the presence of failures and compare the results without functionality degradation.

\subsection{Simulation Setting}

\begin{figure}[thpb]
        \centering
        %\framebox{\parbox{3in}{hhh}}
        \includegraphics[scale=0.35]{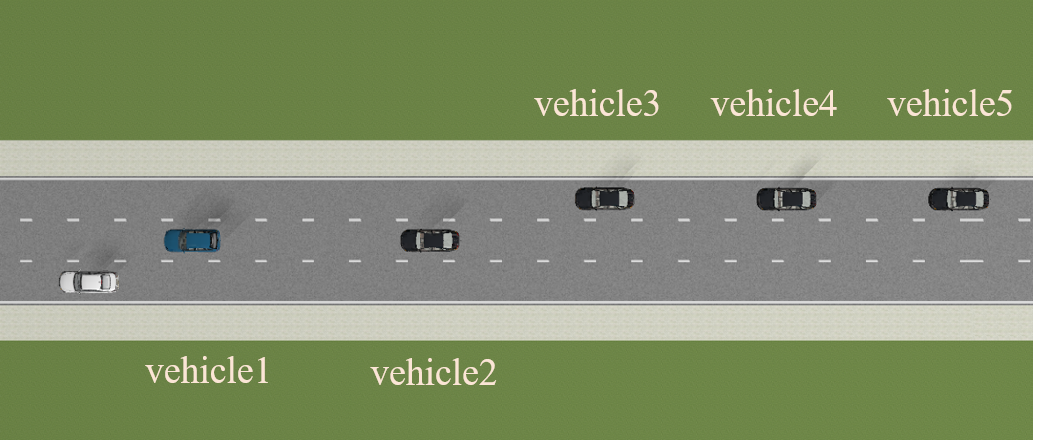}
        \caption{The initial scene of the simulation. The vehicle number remains the same in the whole simulation process.} 
        \label{sim}
      \end{figure}

A simulation platform for a five-vehicle platoon driving in a three-lane highway traffic scenario is established, based on Prescan and Matlab. The initial scene is depicted in Figure~\ref{sim}. As shown in the figure, all the vehicles are driven by the drivers and they are not the members of the platoon at this time. Each vehicle has the identical control system, which integrates different maneuvers in the management layer. The blue vehicle marked as {vehicle1} is the potential leader, the other four black vehicles ranging from {vehicle2} to {vehicle5} are the potential followers, and the white vehicle is the intruding car. Table~\ref{param} details the parameters used in the simulation. 

This paper focuses on the highway scenario, so the normal
driving speed of the platoon is set to 20m/s in the simulation process. The longitudinal controller used in the simulation is a PID controller based on the variable headway time.

      \begin{table}[thpb]
	\centering
	\caption{Parameters employed in the simulation}
	\label{param} 
        \begin{tabular}{cc}
		\hline\hline\noalign{\smallskip}	
		Parameters of CACC Controllers & Value\\
		\hline\hline\noalign{\smallskip}
		variable headway time baseline & $0.5s$\\
                \noalign{\smallskip}\hline\noalign{\smallskip}	
		maximum of headway time & $0.75s$\\
                \noalign{\smallskip}\hline\noalign{\smallskip}	
		minimum of headway time & $0.25s$\\
                \noalign{\smallskip}\hline\noalign{\smallskip}	
		safe distance at a standstill  & $3m$\\
                \hline\hline\noalign{\smallskip}	
		Parameters of Vehicles & Value\\
		\hline\hline\noalign{\smallskip}
		maximum acceleration & $0.30g$\\
                \noalign{\smallskip}\hline\noalign{\smallskip}	
		maximum deceleration & $1.00g$\\
                \noalign{\smallskip}\hline\noalign{\smallskip}	
		initial velocity  & $20 m/s$\\
                \noalign{\smallskip}\hline\noalign{\smallskip}
	\end{tabular}
\end{table}

\subsection{Overall Platoon Performance of Integrated maneuvers}
\begin{figure*}[thpb]
        \centering
        %\framebox{\parbox{3in}{hhh}}
        \includegraphics[scale=0.5]{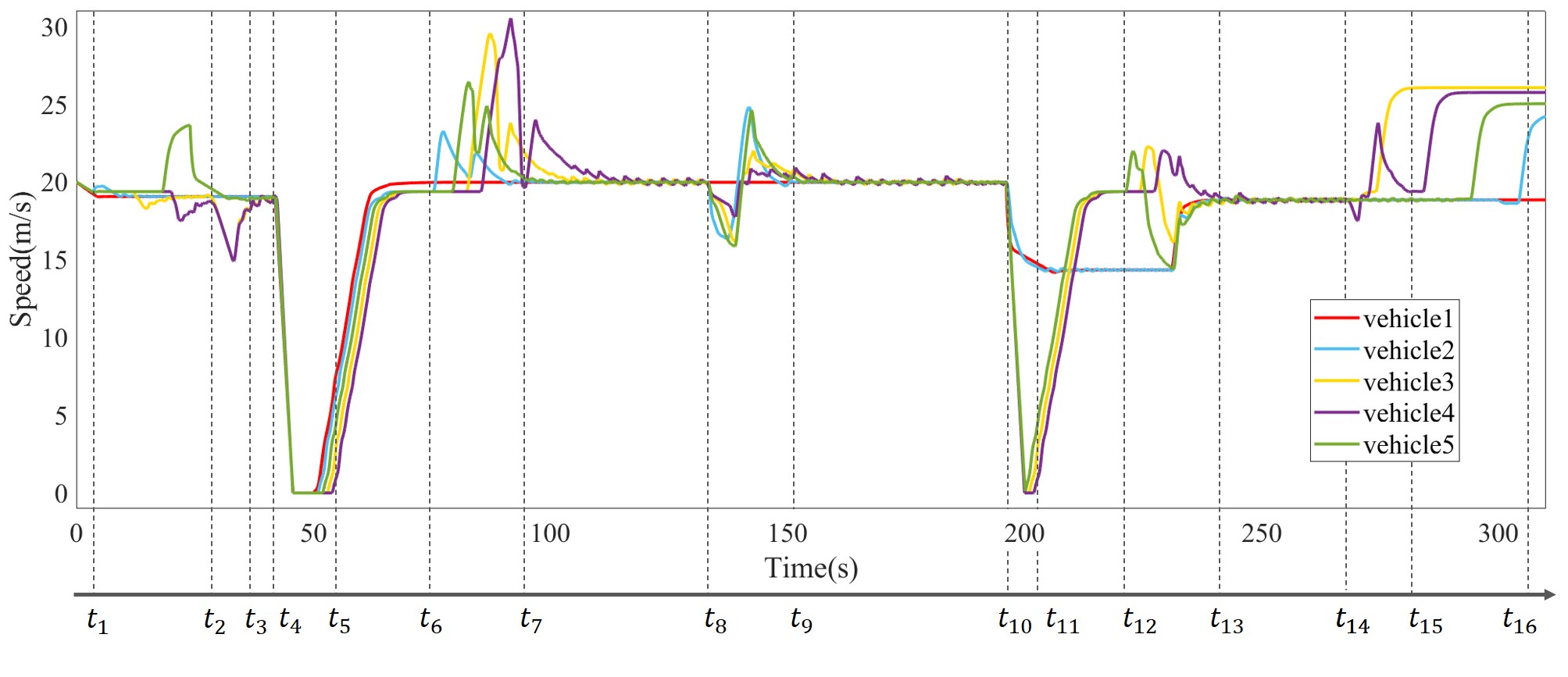}
        \caption{Speed trajectories of platoon vehicles. The maneuvers mentioned before occur in the whole process.} 
        \label{speed}
      \end{figure*}
      
In order to examine the behavior of the proposed framework, an integrated simulation is performed with the sequence of seven platoon maneuvers: Join Tail, Join Middle, AEB Head, Cut In, AEB middle, Leave Middle and Leave Tail. The simulation result could be seen in Figure~\ref{speed}. The whole process could be divided into the seven following parts and the speed trajectories prove the behavior of the maneuver management framework.  

\subsubsection{${t_1} \sim {t_2}$ } Join Tail.  According to Algorithm~\ref{JTFREE}, {vehicle2}, {vehicle3}, and {vehicle4} perform Join Tail maneuver one after one where there is a small adaption for the vehicle speed. And {vehicle5} will accelerate to prepare for joining middle.

\subsubsection{${t_2} \sim {t_3}$ } Join Middle. {Vehicle5} joins the platoon before {vehicle3}. {Vehicle3} decelerates to increases the distance, and {vehicle4} also decelerates to follow {vehicle3}. After reaching the safe distance, {vehicle3} returns to the initial speed.

\subsubsection{${t_4} \sim {t_5}$ } AEB Head. In this period, an intruding car suddenly cuts in before the lead vehicle and the TTC condition is satisfied, so the AEB system of the leader activates and the whole platoon performs emergency braking. After the intruding car is gone, the platoon will be restarted by the drivers and followed by the Join maneuver in the $t_6 \sim t_7$ period.

\subsubsection{${t_8} \sim {t_9}$ } Cut In. In this period, an intruding car cuts in before {vehicle2} while the distance is not small enough so the AEB system is not activated. Then {vehicle2} and the following vehicles select the ACC controller to increase the intervehicle distance. After the car cuts out, the distance decreases.

\subsubsection{${t_{10}} \sim {t_{11}}$ } AEB Middle. An intruding car cuts in before {vehicle5} and the TTC condition is satisfied, so {vehicle5} and the following vehicles perform emergency braking, while the leader will decelerate to wait for them. After the car is gone, they will be restarted by the driver and followed by the Join maneuver in the $t_{12} \sim t_{13}$ period.

\subsubsection{${t_{14}} \sim {t_{15}}$ } Leave Middle. When {vehicle3} receives the leave instruction, {vehicle4} needs to decelerate to increase the gap between them and then {vehicle3} changes lane to leave the platoon and it will be controlled by the driver.

\subsubsection{${t_{15}} \sim {t_{16}}$ } Leave Tail. In this period, {vehicle4}, {vehicle5} and {vehicle2} change lanes and leave the platoon one by one.

\subsection{Specific Analysis of Platoon Performance} 

The simulation results of two typical maneuvers (Join Tail and AEB Head) are shown in this part, including the sequence of snapshots, speed trajectories, and intervehicle distance curves.

\subsubsection{Join Tail Result}
\begin{figure}[H]
	\centering
	\begin{subfigure}{0.49\linewidth}
		\centering
		\includegraphics[width=0.9\linewidth]{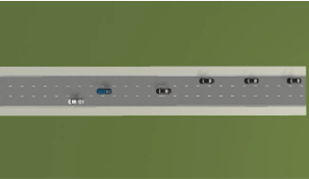}
		\caption{before Join Tail}
                \label{snapjt1}
	\end{subfigure}
	\begin{subfigure}{0.49\linewidth}
		\centering
		\includegraphics[width=0.9\linewidth]{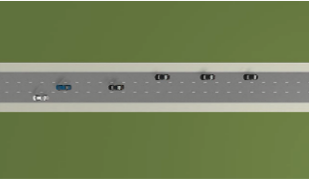}
		\caption{Join Tail completes}
                \label{snapjt2}
	\end{subfigure}
        \caption{A sequence of snapshots of the Join Tail maneuver.}
        \label{snapjt}
\end{figure}

The simulation result of Join Tail is illustrated as a sequence of snapshots in Fig.~\ref{snapjt}. In Fig.~\ref{snapjt1}, vehicle2 receives the instruction to join the platoon after vehicle1, then vehicle2 accelerates,decreasing the intervehicle distance between vehicle1. After vehicle2 reaches the required distance, it becomes the follower in the platoon. This is shown in Fig.~\ref{snapjt2}.

\begin{figure}[thbp]
        \centering
        \begin{subfigure}[b]{0.4\textwidth}
               \centering
               \includegraphics[width=\textwidth]{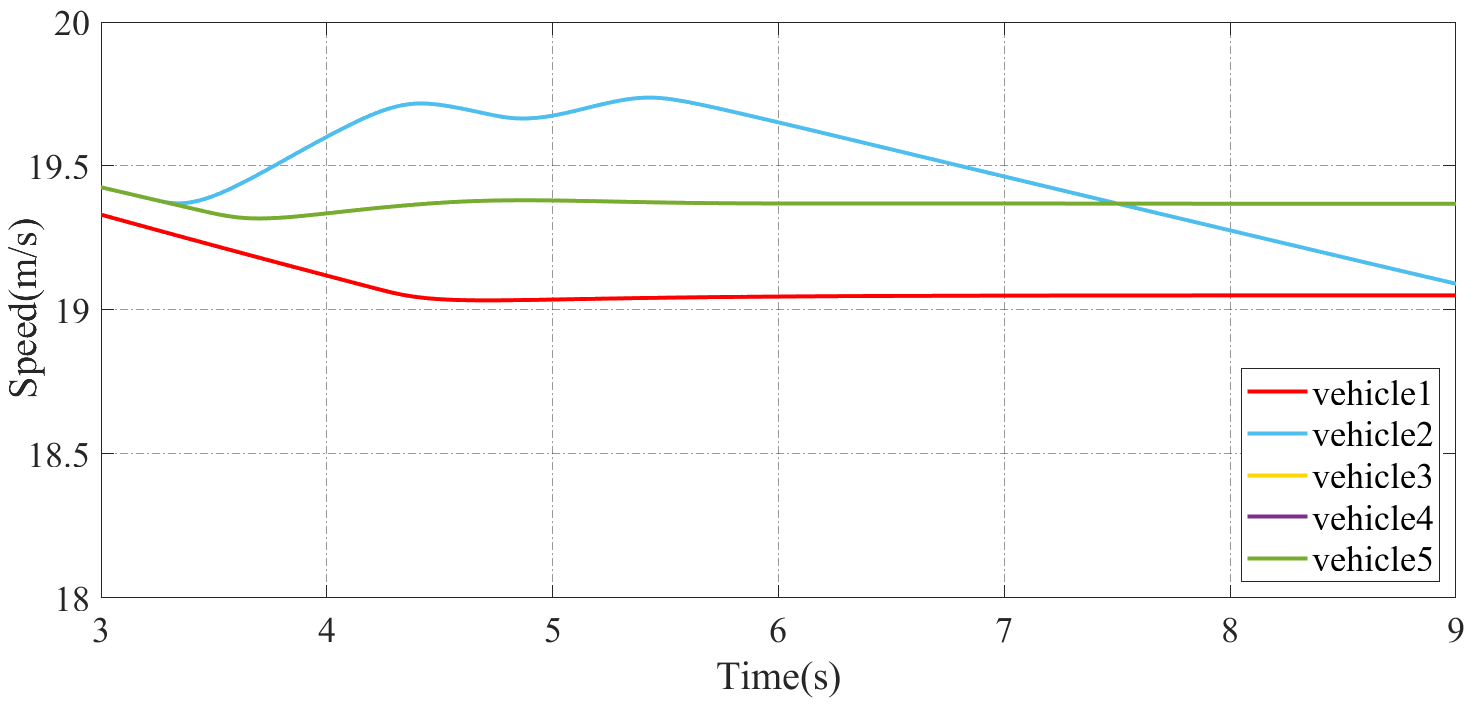}
                \caption{Speed trajectories}
                \label{jtv}
        \end{subfigure}
        \begin{subfigure}[b]{0.4\textwidth}
                \centering
                \includegraphics[width=\textwidth]{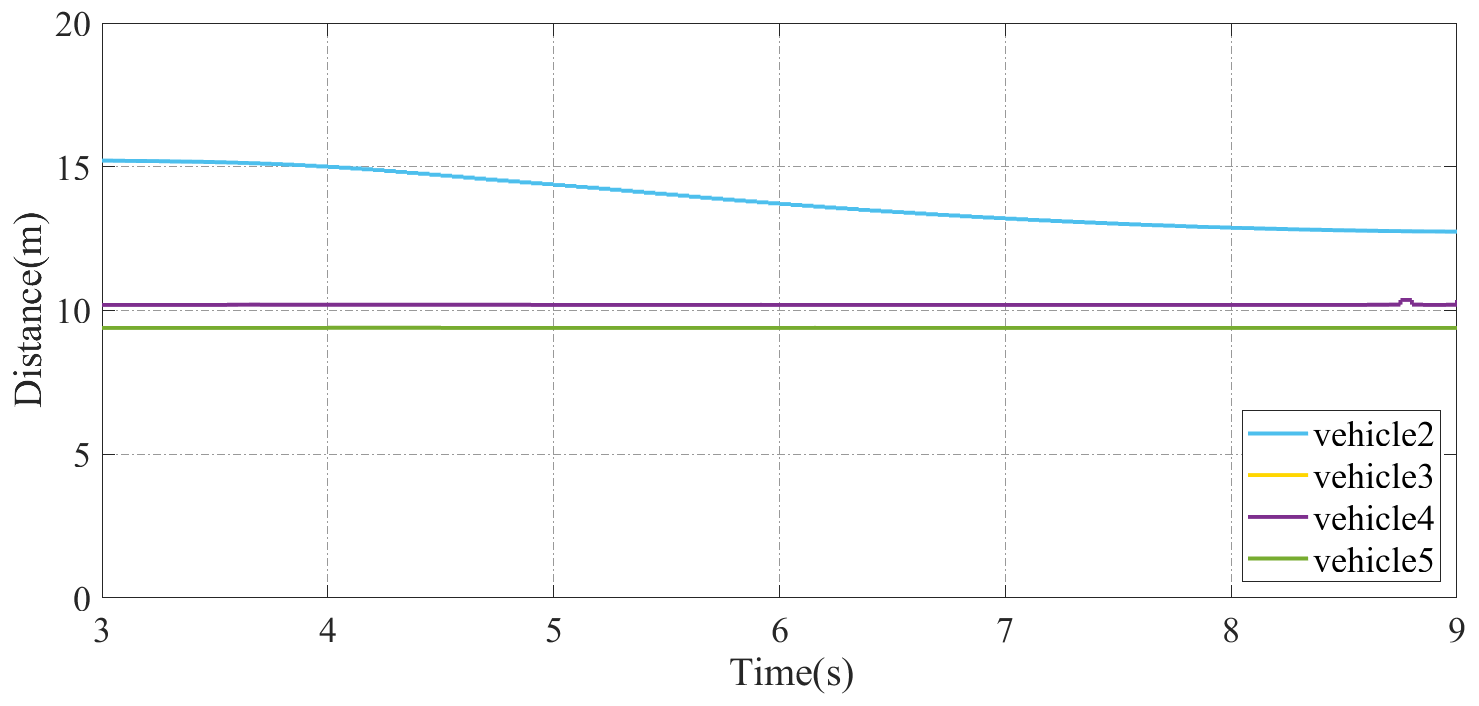}
                \caption{Intervehicle distance}
                \label{jtdis}
        \end{subfigure}
        \caption{Simulation result of Join Tail}
    \end{figure}

Besides, Fig.~\ref{jtv} shows the speed trajectories and Fig.~\ref{jtdis} shows the intervehicle distance detected by the radar. These curves verify the results of the simulation numerically. As shown in Fig.~\ref{jtv}, when receiving the join instruction, vehicle2 accelerates to reach the required intervehicle distance. {Vehicle4} also decelerates to follow {vehicle3}. Fig.~\ref{jtdis} further validates this result. The distance curve of vehicle2 falls, proving that it accelerates to join the platoon from tail. The simulation results prove that the proposed framework and management strategy could achieve the ability of the Join Tail maneuver.
   
\subsubsection{AEB Head Result}
\begin{figure}[htbp]
	\centering
	\begin{subfigure}{0.49\linewidth}
		\centering
		\includegraphics[width=0.9\linewidth]{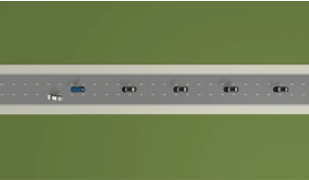}
		\caption{before AEB Head}
                \label{snapAEB1}
	\end{subfigure}
	\begin{subfigure}{0.49\linewidth}
		\centering
		\includegraphics[width=0.9\linewidth]{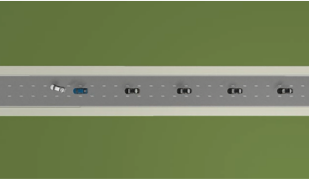}
		\caption{perform AEB}
                \label{snapAEB2}
	\end{subfigure}
	
	\begin{subfigure}{0.49\linewidth}
		\centering
		\includegraphics[width=0.9\linewidth]{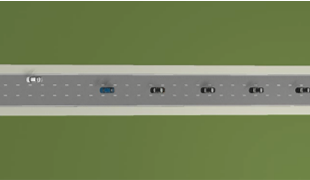}
		\caption{the intruding car cuts out}
                \label{snapAEB3}
	\end{subfigure}
	\begin{subfigure}{0.49\linewidth}
		\centering
		\includegraphics[width=0.9\linewidth]{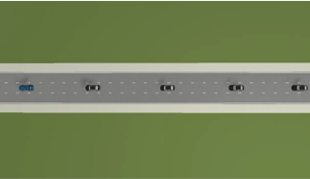}
		\caption{platoon starts sequentially}
                \label{snapAEB4}
	\end{subfigure}
        \caption{A sequence of snapshots of the AEB Head maneuver.}
        \label{snapAEB}
\end{figure}

The simulation result of AEB Head is illustrated as a sequence of snapshots in Fig.~\ref{snapAEB}. In Fig.~\ref{snapAEB1}, the white intruding car begins to change lane suddenly and when it drives into the lane of the platoon, it is very close to the leader. As a result, the TTC condition is satisfied and the AEB system is activated as shown in Fig.~\ref{snapAEB2}. In Fig.~\ref{snapAEB3}, the intruding car cuts out. After that, the vehicles will be restarted from a standstill by the drivers.

\begin{figure}[htbp]
        \centering
        \begin{subfigure}[b]{0.4\textwidth}
               \centering
               \includegraphics[width=\textwidth]{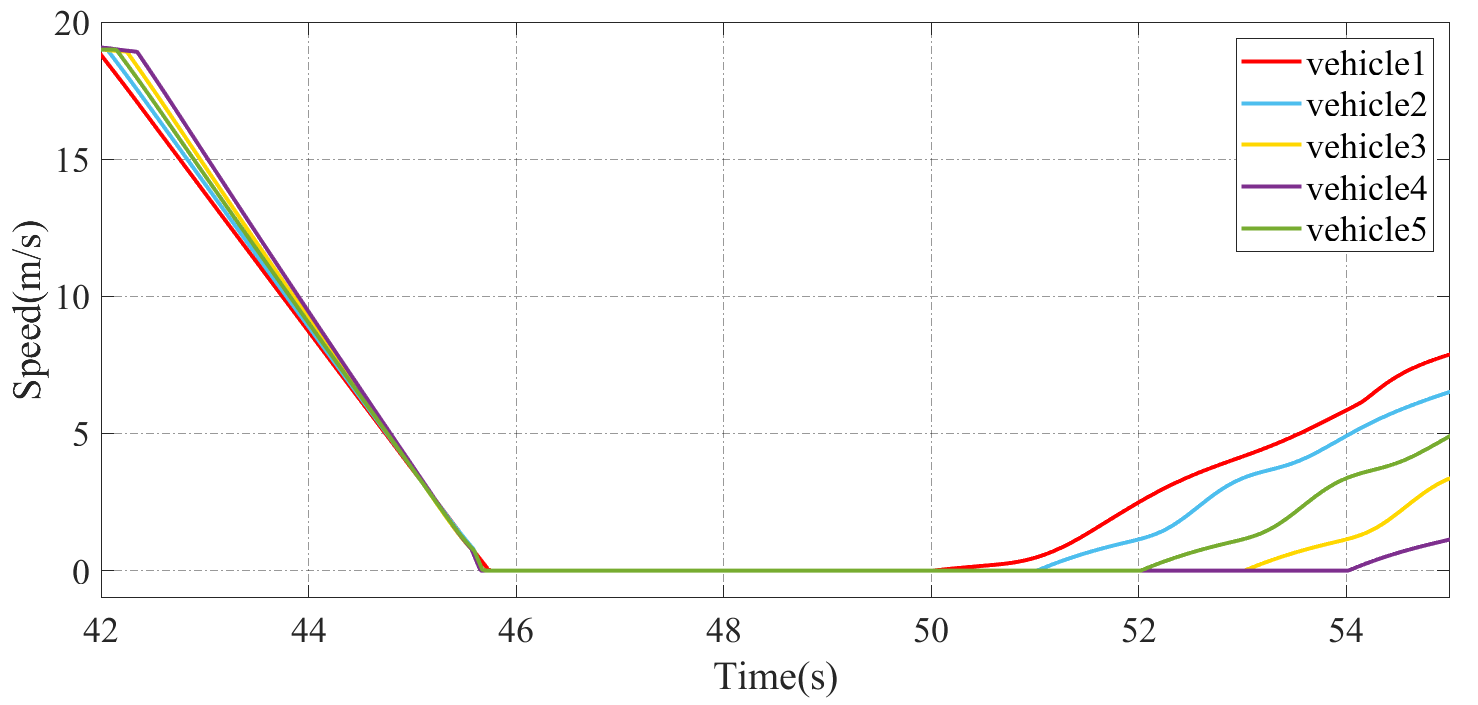}
                \caption{Speed trajectories}
                \label{AEBv}
        \end{subfigure}
        \begin{subfigure}[b]{0.4\textwidth}
                \centering
                \includegraphics[width=\textwidth]{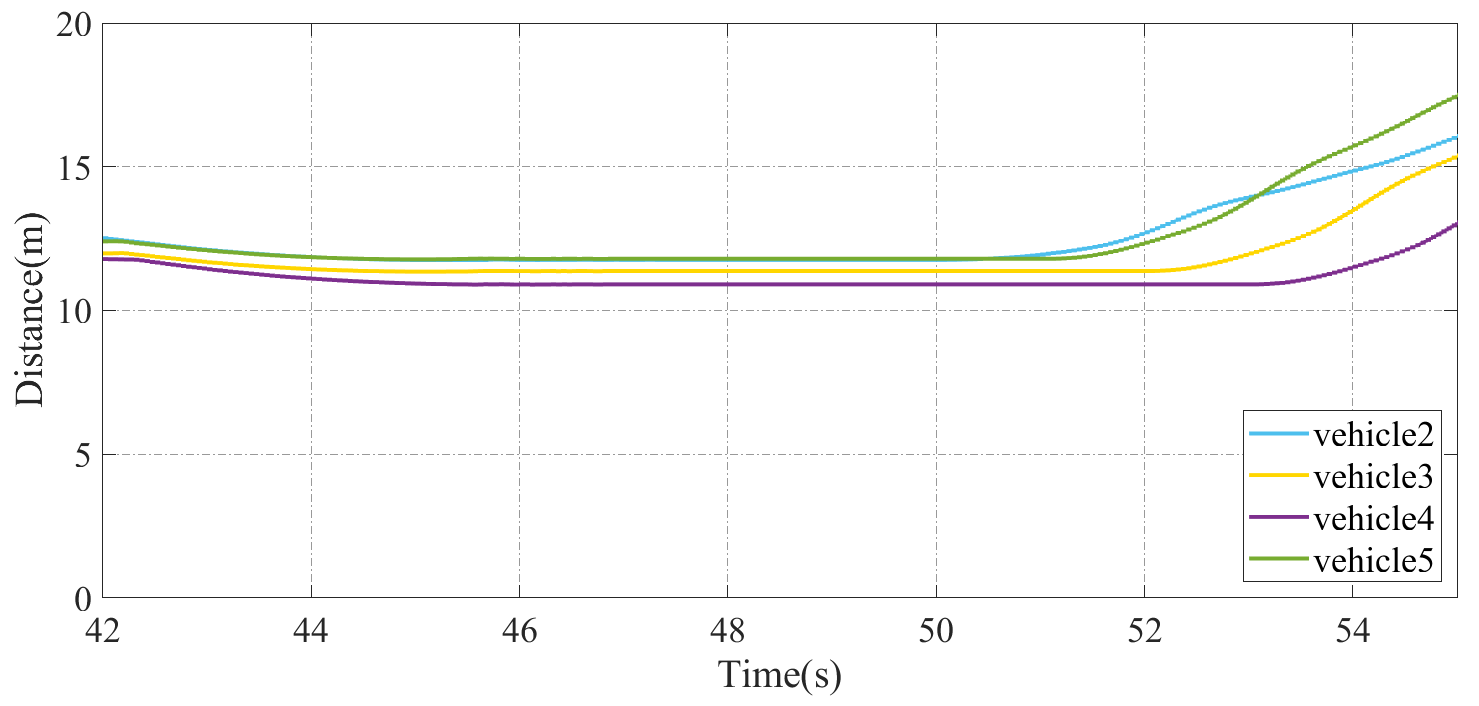}
                \caption{Intervehicle distance}
                \label{AEBdis}
        \end{subfigure}
        \caption{Simulation result of AEB Head}
    \end{figure}

As can be seen from the simulation result of Fig.~\ref{AEBv}, the vehicles in the platoon decelerate nearly at the same time when the intruding car cuts in suddenly. After decelerating to a standstill, the vehicles wait until they are restarted by the drivers one by one. Fig.~\ref{AEBdis} shows the intervehicle distance detected by the radar and further validates this result. The distances between each vehicle in the platoon are all more than $10m$, which proves that the management strategy of AEB Head maneuver could keep the platoon safe when the emergency braking happens. The simulation results validate the proposed framework and management strategy could achieve the ability of the AEB Head maneuver. Besides, the security of the platoon is ensured at the same time for collision avoidance.

\subsection{Fault-Tolerant Mechanism Performance}

This part shows the simulation results of the platoon control system with and without the fault-tolerant mechanism in the presence of V2V and radar failures separately. Through the comparison, the key importance of the functionality degradation is verified.

\subsubsection{V2V Failure Result}

\begin{figure*}[htbp]
	\centering
	\begin{subfigure}{0.49\linewidth}
		\centering
		\includegraphics[width=0.9\linewidth]{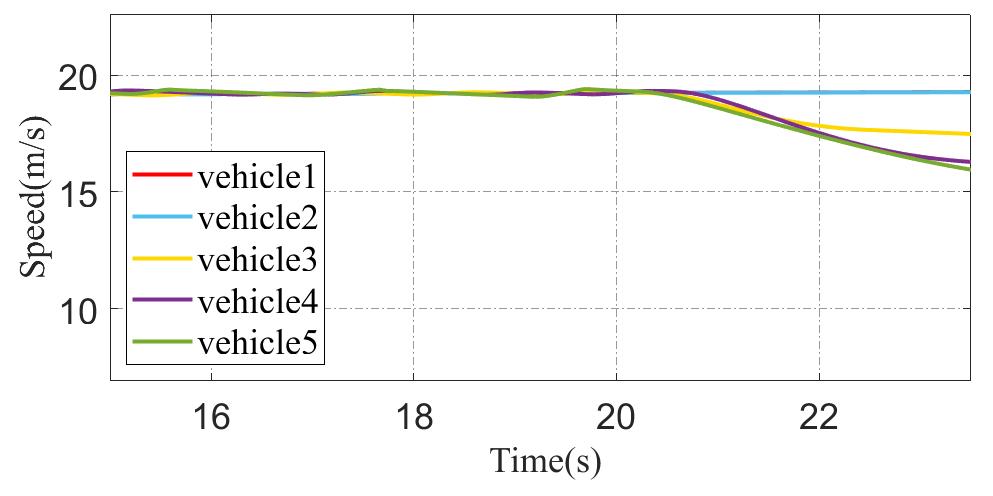}
		\caption{Speed trajectories with degradation}
                \label{velV2Vyes}
	\end{subfigure}
	\begin{subfigure}{0.49\linewidth}
		\centering
		\includegraphics[width=0.9\linewidth]{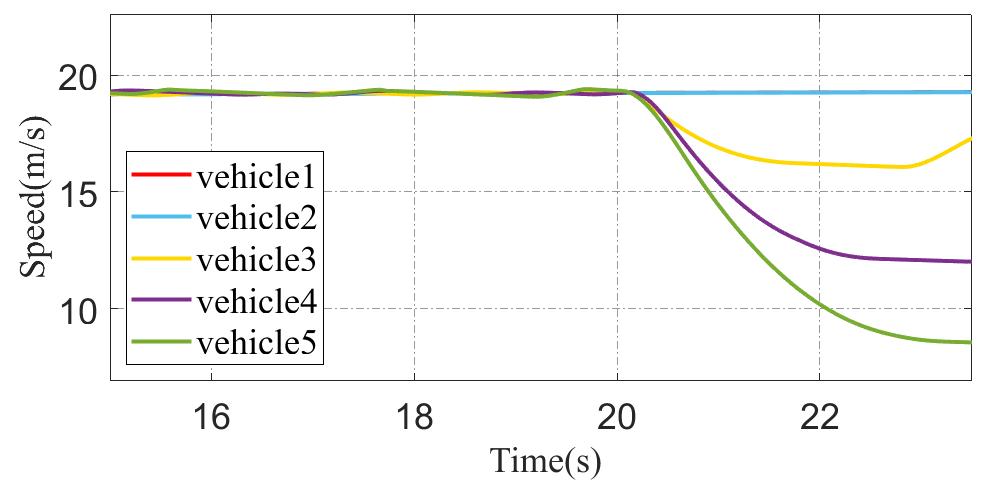}
		\caption{Speed trajectories without  degradation}
                \label{velV2Vno}
	\end{subfigure}
	\begin{subfigure}[b]{0.49\linewidth}
		\centering
		\includegraphics[width=0.9\linewidth]{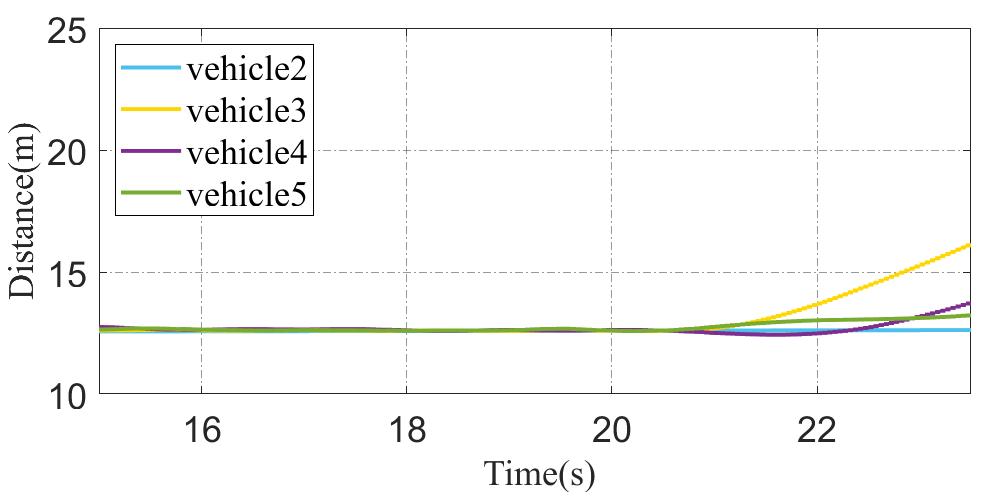}
		\caption{Intervehicle distance with degradation}
                \label{disV2Vyes}
	\end{subfigure}
	\begin{subfigure}[b]{0.49\linewidth}
		\centering
		\includegraphics[width=0.9\linewidth]{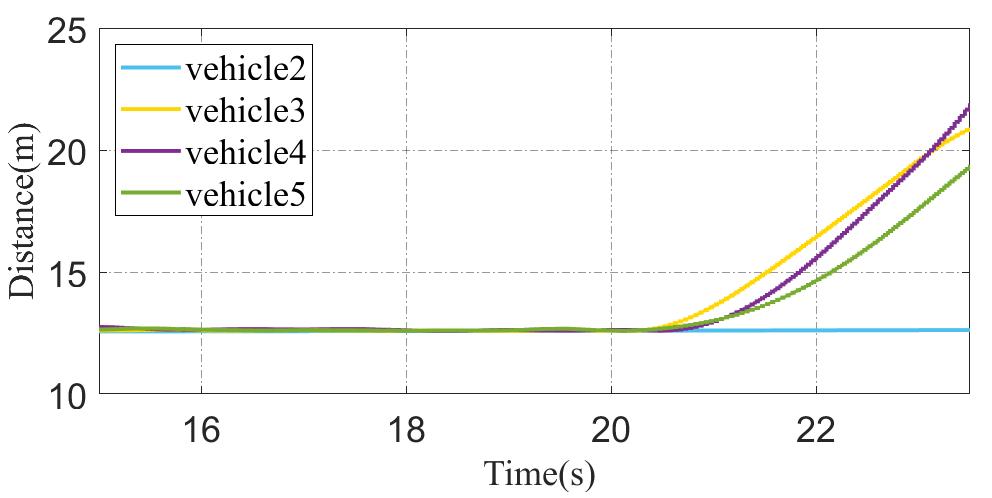}
		\caption{Intervehicle distance without degradation}
                \label{disV2Vno}
	\end{subfigure}
        \caption{Simulation results when the V2V device fails at 20s.}
        \label{V2V}
\end{figure*}

\begin{figure*}[htbp]
	\centering
	\begin{subfigure}{0.49\linewidth}
		\centering
		\includegraphics[width=0.9\linewidth]{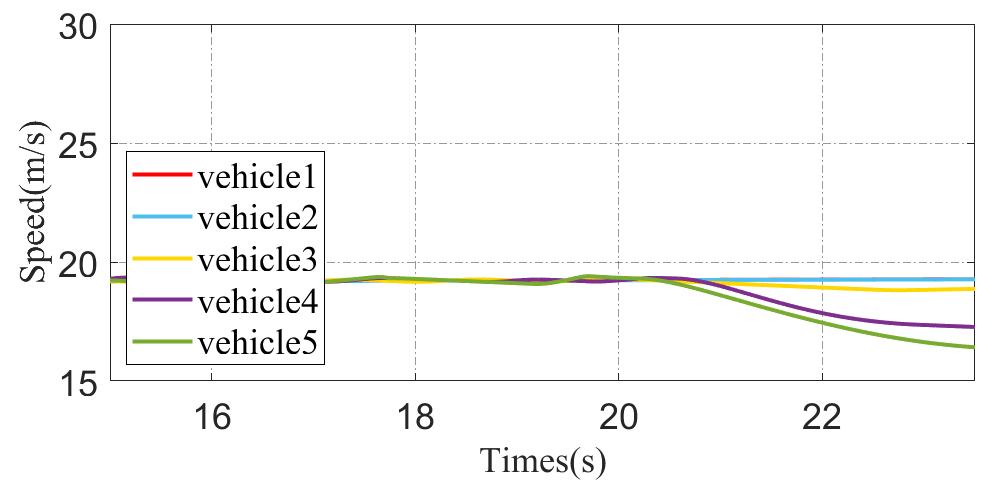}
		\caption{Speed trajectories with degradation}
                \label{velradaryes}
	\end{subfigure}
	\begin{subfigure}{0.49\linewidth}
		\centering
		\includegraphics[width=0.9\linewidth]{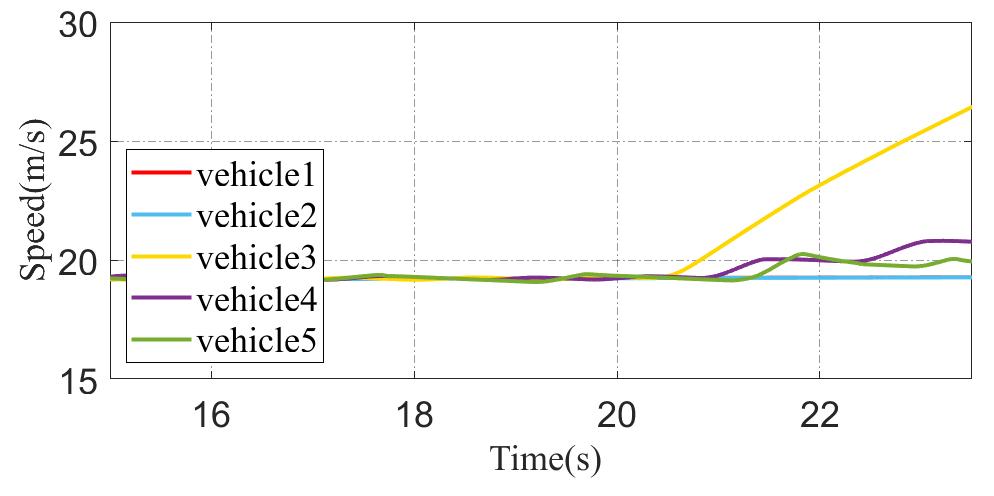}
		\caption{Speed trajectories without degradation}
                \label{velradarno}
	\end{subfigure}
	\begin{subfigure}[b]{0.49\linewidth}
		\centering
		\includegraphics[width=0.9\linewidth]{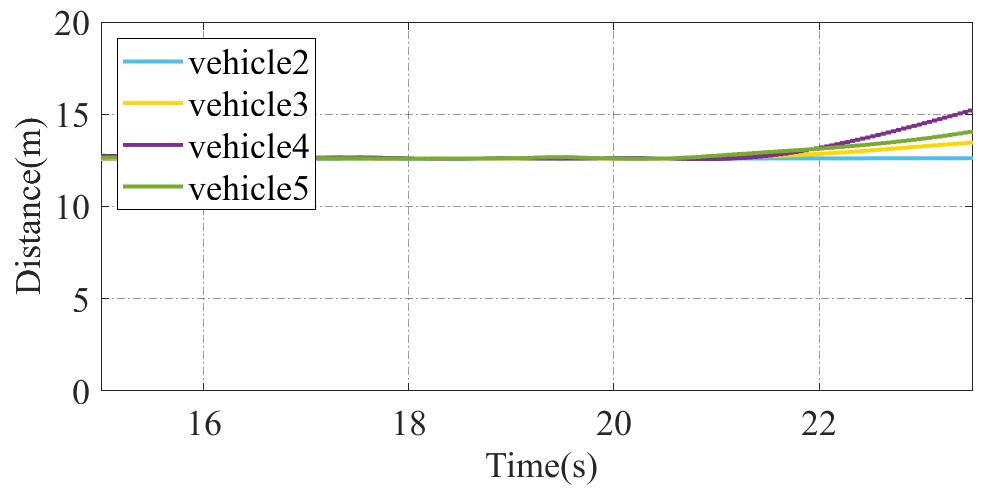}
		\caption{Intervehicle distance with degradation}
                \label{disradaryes}
	\end{subfigure}
	\begin{subfigure}[b]{0.49\linewidth}
		\centering
		\includegraphics[width=0.9\linewidth]{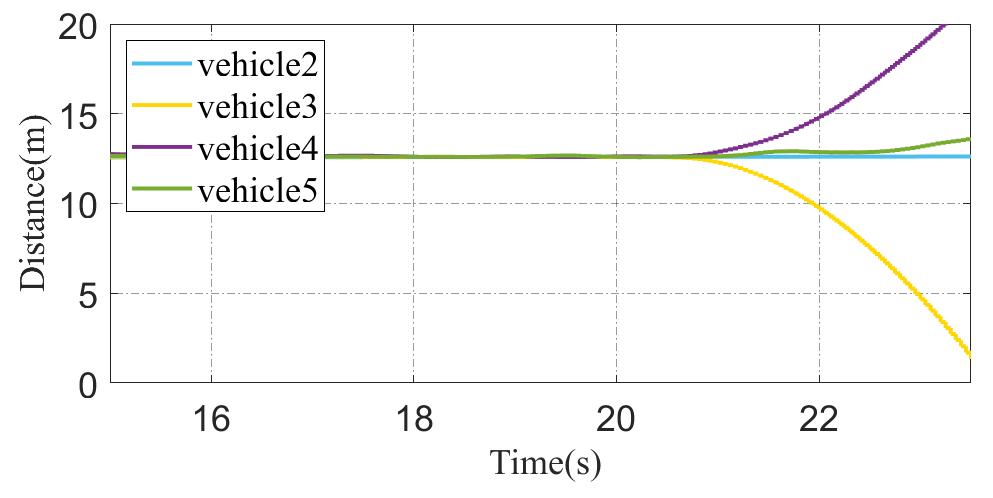}
		\caption{Intervehicle distance without degradation}
                \label{disradarno}
	\end{subfigure}
        \caption{Simulation results when the radar fails at 20s.}
        \label{radar}
\end{figure*}

Fig.~\ref{V2V} show the simulation results when the V2V device of vehicle3 fails at $20s$. The left column is the results with the fault-tolerant mechanism while the right column not. 

The vehicles are in steady platooning state before the V2V failure. The travelling speed is about $20 m/s$, and the intervehicle distance is about $13 m$. When the V2V device fails, the fault detection module in the platoon vehicles will detect the failure and then the state is switched to the Hardware Failures maneuver. According to the fault-tolerant mechanism, vehicle3, vehicle4 and vehicle5 will degrade the longitudinal controller to the ACC controller while vehicle1 and vehicle2 keep using the CACC controller. As can be seen from Fig.~\ref{velV2Vyes} and Fig.~\ref{disV2Vyes}, the speed of vehicle3 and the following vehicles is reduced after $20s$, and the intervehicle distance of the three vehicles increases. While vehicle1 and vehicle2 can still keep steady platooning. The simulation results show that the degradation of controller could maintain the temporary safty of the platoon system before the takeover when the V2V device on one vehicle fails. 

Compared with the results with degradation, Fig.~\ref{velV2Vno} and Fig.~\ref{disV2Vno} show the results without degradation. When the V2V device fails, the communication data from vehicle3 all turns to zero. The nominal functionality of the CACC controller could not maintain. According to our logic for designing the CACC controller, the actual distance that vehicle3 and the following vehicles can acquire will be much smaller than the desired distance. As a result, the three vehiles will decelerate rapidly, which is highly likely to cause rear-end collision.

\subsubsection{Radar Failure Result}

Fig.~\ref{radar} show the simulation results when the radar of vehicle3 fails at $20s$. The left column is the results with the fault-tolerant mechanism while the right column not. 

When the V2V device fails, the faulty vehicle will broadcast the failure message and then the states of the platoon vehicles are switched to the Hardware Failures maneuver. According to the fault-tolerant mechanism, vehicle3, the faulty vehicle will degrade to the CC controller due to the lack of distance data. The following vehicles, vehicle4 and vehicle5 will use the ACC controller. In Fig.~\ref{velradaryes} and Fig.~\ref{disradaryes} we can see that vehicle3 travels at a constant speed which is a little lower than $20m/s$, and the speeds of following vehicles are reduced after the radar fails. The intervehicle distances of the faulty vehicle and the following vehicles all increase.

Compared with the results with degradation, Fig.~\ref{velradarno} and Fig.~\ref{disradarno} show the results without degradation. When the radar fails, the distance detected by radar becomes very far. The actual distance acquired by vehicle3 will be much larger than the desired distance. As a result, vehicle3 will accelerate rapidly, and collides with vehicle2 at $23.5s$. The simulation results show that the platoon system with faulty-tolerant mechanism could maintain the temporary safty before the driver takes over when the radar on one vehicle fails.

\section{REAL MICRO-VEHICLE EXPERIMENT}

In this section, the real-world experiment is conducted on an outdoor road using three micro-vehicles, and the results of the experiment verified the effectiveness and real-time performance of the proposed framework.

\subsection{Experiment Setting}

\begin{figure}[thpb]
        \centering
        %\framebox{\parbox{3in}{hhh}}
        \includegraphics[scale=0.45]{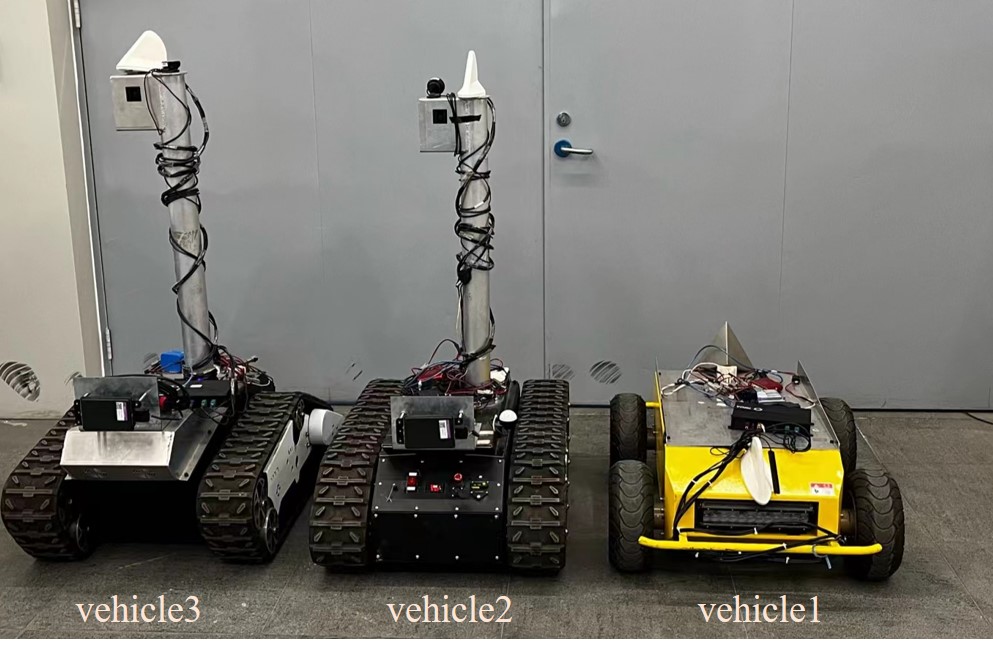}
        \caption{The micro-vehicles for the experiment.} 
        \label{real}
     \end{figure}

All three micro-vehicle are made by JROBOT, as shown in Fig.~\ref{real}. Vehicle1 is a wheeled model, WARTHOG01 and acts as the leader in the platoon after initializing. Vehicle2 and vehicle3 are the tracked models Komodo, both of which will become the followers after the Join Tail Maneuver. 

\begin{figure}[thpb]
	\centering
	\begin{subfigure}{0.49\linewidth}
		\centering
		\includegraphics[width=0.9\linewidth]{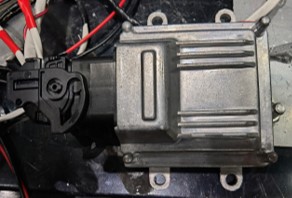}
		\caption{Vehicle control unit (VCU)}
                \label{vcu}
	\end{subfigure}
	\begin{subfigure}{0.49\linewidth}
		\centering
		\includegraphics[width=0.9\linewidth]{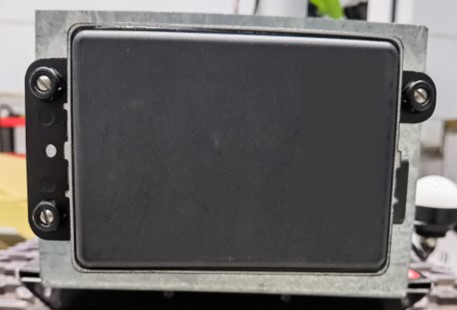}
		\caption{V2V device}
                \label{v2x}
	\end{subfigure}
	
	\begin{subfigure}{0.49\linewidth}
		\centering
		\includegraphics[width=0.9\linewidth]{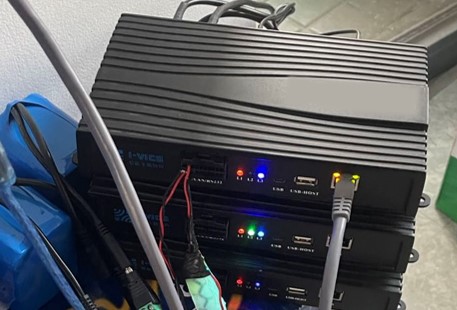}
		\caption{Millimeter wave radar}
                \label{millradar}
	\end{subfigure}
	\begin{subfigure}{0.49\linewidth}
		\centering
		\includegraphics[width=0.9\linewidth]{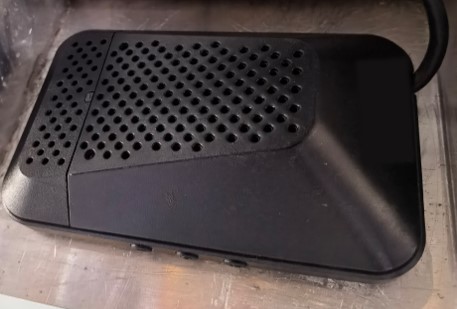}
		\caption{Monocular camera}
                \label{camera}
	\end{subfigure}
        \caption{The hardwares equipped on the micro-vehicle.}
        \label{real_hardware}
\end{figure}

All hardwares equipped on the micro-vehicles are shown in Fig.~\ref{real_hardware}. The first one is a vehicle control unit(VCU), which serves as the core calculating function, with the programs of the communication layer, the management layer and the control layer running on it. Fig.~\ref{v2x} shows the V2V device OBUYZM9 made by NEBULA-LINK is adopted in this experiment. This module provides V2V communication through 4G/LTE. In addition to the devices on the micro-vehicles, there is another one used to simulate the cloud layer to send instructions. Fig.~\ref{millradar} shows the millimeter wave radar, Delphi ESR and Fig.~\ref{camera} shows the monocular camera, Mobileye 630.

The experiment in this paper is carried out on the four-lane road segment in Tsinghua University and only one lane was utilized in this experiment. The chosen segment contains a straight section and curved section. The cloud layer sent the Join Tail instruction to the vehicle2 and vehile3 in order. After joining in the platoon, the vehicles travel steadily through the curved segment.

\subsection{Experimenal Validation Results}

\begin{figure*}[thbp]
        \centering
        \begin{subfigure}[b]{0.22\textwidth}
               \centering
               \includegraphics[width=\textwidth]{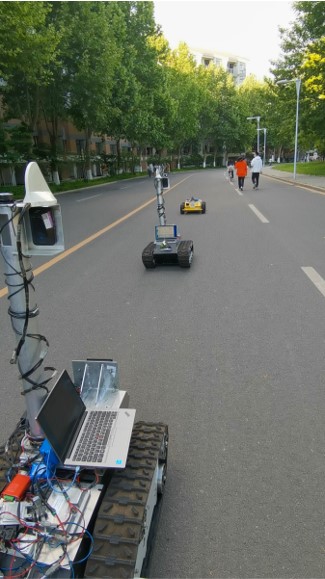}
                \caption{Vehicle2 started joining.}
                \label{real1}
        \end{subfigure}
        \begin{subfigure}[b]{0.22\textwidth}
                \centering
                \includegraphics[width=\textwidth]{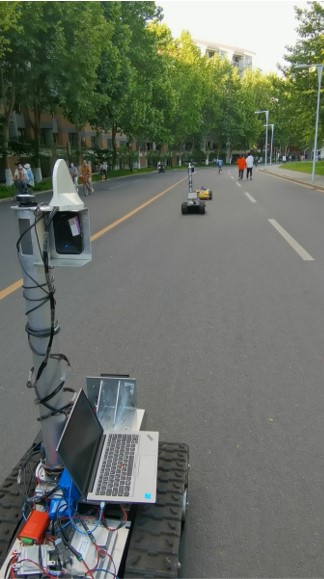}
                \caption{Vehicle2 finished joining.}
                \label{real2}
        \end{subfigure}
        \begin{subfigure}[b]{0.22\textwidth}
                \centering
                \includegraphics[width=\textwidth]{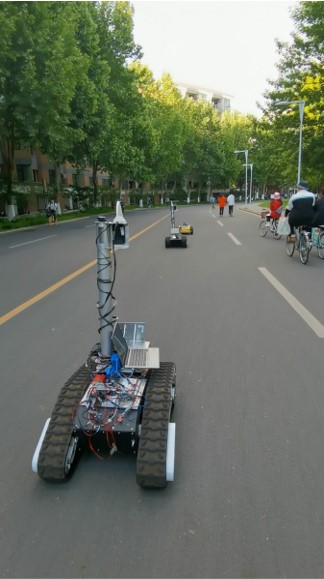}
                \caption{Vehicle3 started joining.}
                \label{real3}
        \end{subfigure}
        \begin{subfigure}[b]{0.22\textwidth}
                \centering
                \includegraphics[width=\textwidth]{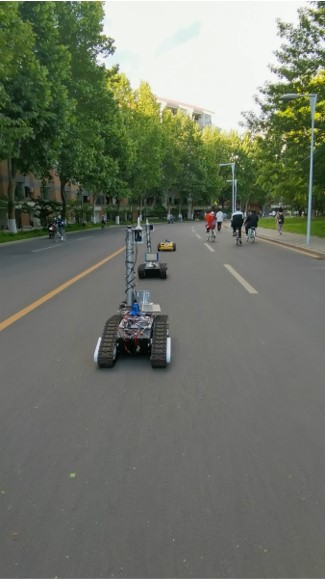}
                \caption{Vehicle3 finished joining.}
                \label{real4}
        \end{subfigure}
        \caption{Snapshots of the field experiment in Join Tail maneuver.}
        \label{real_video}
    \end{figure*}

The snapshots of the micro-vehicle experiment are shown in Fig.~\ref{real_video}. On stage 1 in Fig.~\ref{real1}, the cloud layer sends the joining instruction and the joining position to vehicle2. All participating vehicles transition to Join Tail maneuver when receiving the message. Vehicle2 adopts Algorithm~\ref{JTFREE}, accelerating to catch up vehicle1. On state 2 in Fig.~\ref{real2}, the distance between vehicle1 and vehicle2 reaches the set value, then vehicle2 switches to steady platooning. Vehicle1 updates the information of the platoon according to Algorithm~\ref{JTLEAD}. On state 3 in Fig.~\ref{real3}, the cloud layer sends the join instruction again and vehicle3 accelerates in Join Tail maneuver. On stage 4 in Fig.~\ref{real4}, vehicle3 finished joining and the three-vehicle platoon travels steadily through the rest of the segment.

    \begin{figure}[H]
        \centering
        %\framebox{\parbox{3in}{hhh}}
        \includegraphics[scale=0.23]{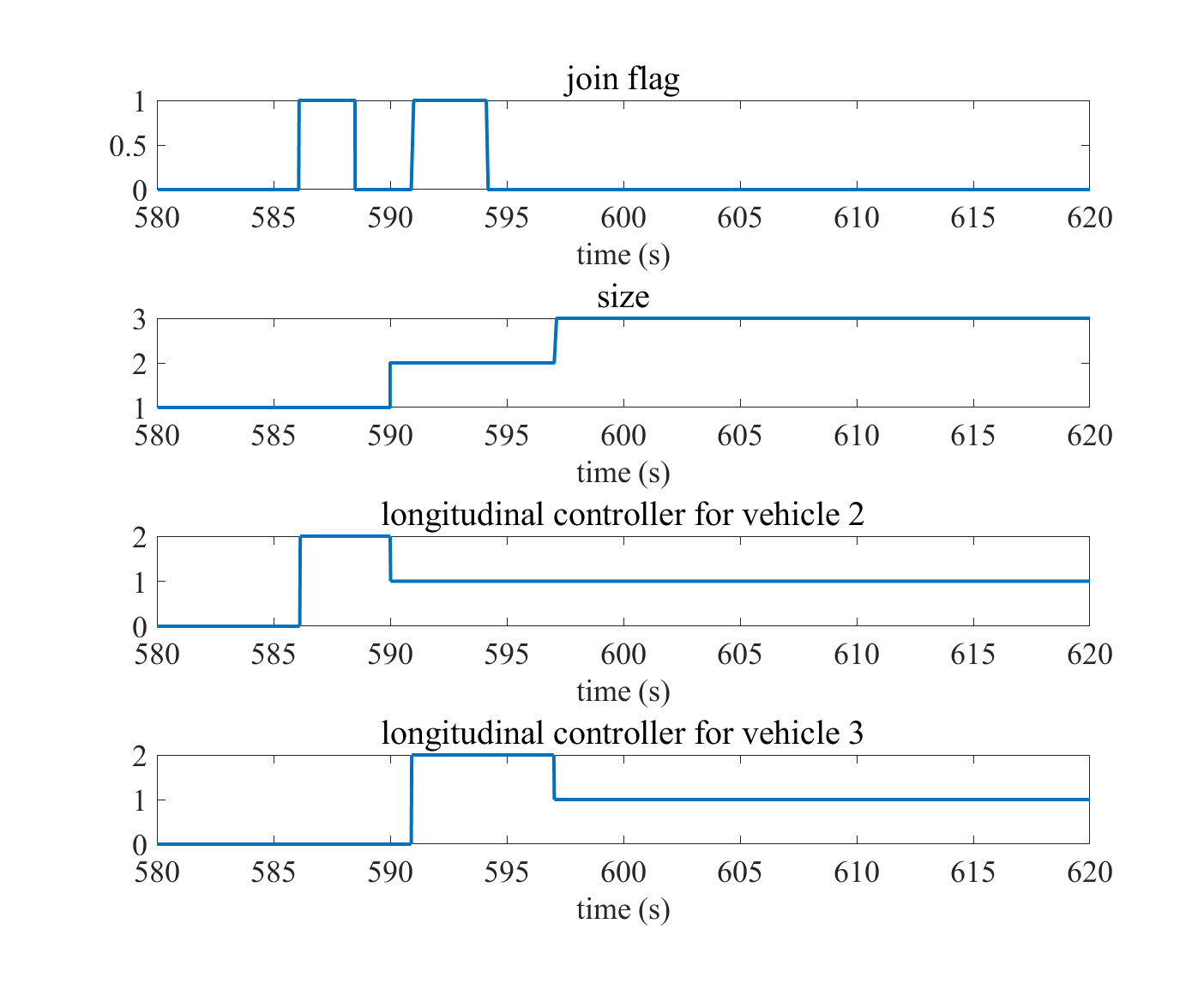}
        \caption{Indicating data in the process of the experiment, including: the joining flag sent by the cloud layer, the size of the platoon updated by the leader and the longitudinal controller adopted by vehicle2 and vehicle3.} 
        \label{real_flag}
     \end{figure}

     \begin{figure}[H]
        \centering
        %\framebox{\parbox{3in}{hhh}}
        \includegraphics[scale=0.23]{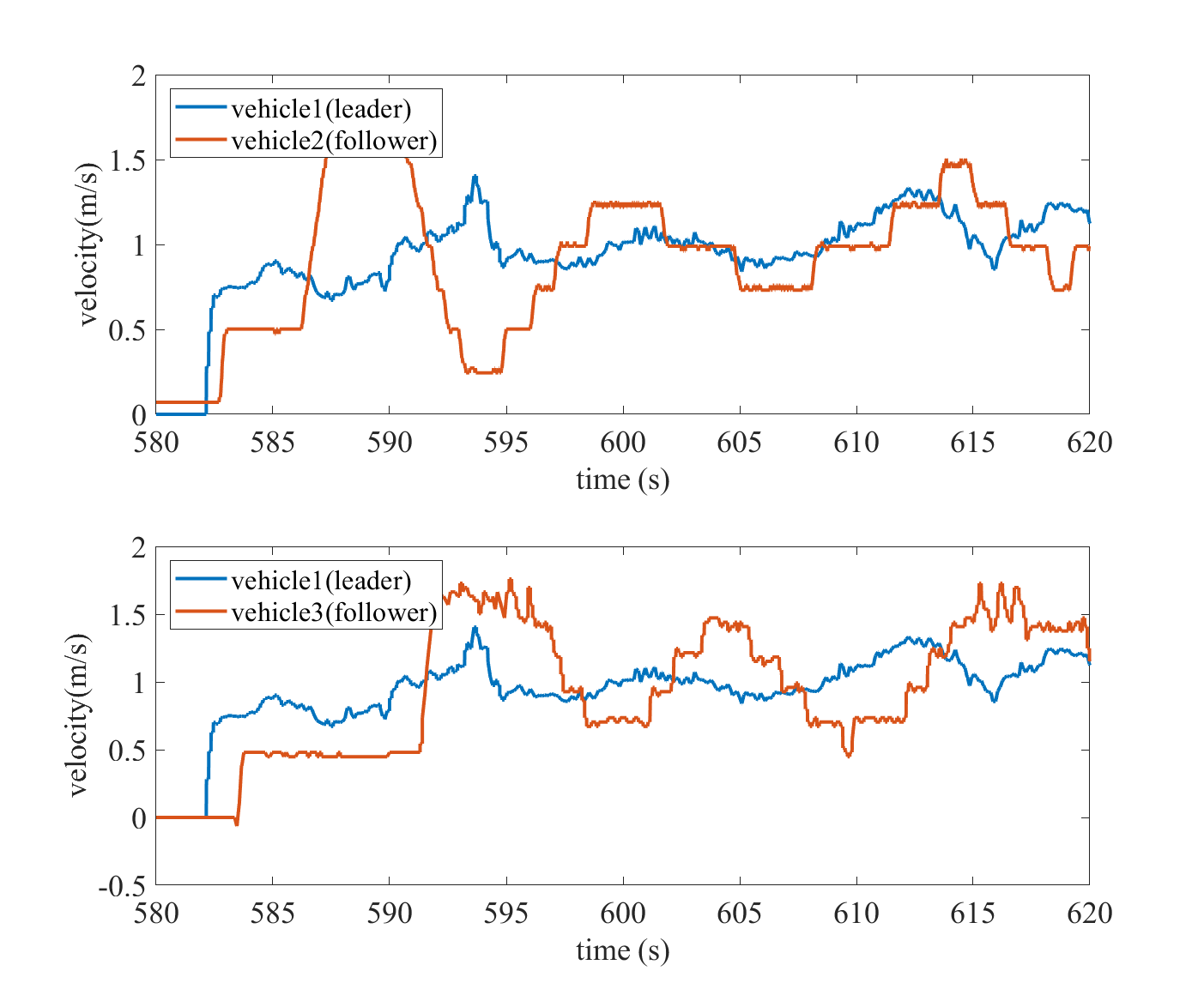}
        \caption{Speed trajectories of vehicles in the process of the experiment.} 
        \label{real_vel}
     \end{figure}
     
As can be seen in Fig.~\ref{real_flag} and Fig.~\ref{real_vel}, the validity of the proposed framework is further demonstrated by the recorded data during the field experiment. Fig.~\ref{real_flag} mainly shows the indicating data during the process. The joining flag is sent twice at 586s and 591s by the cloud layer. After receiving the flag, vehicle2 switches to the ACC controller, which is represented by 2 in the figure. When finishing joining, the controller transitions to the CACC controller which is represented by 1 and the leader update the size of the platoon to 2. The same process happens when the cloud layer sends the joining instruction a second time and updates the size to 3. Fig.~\ref{real_vel} shows the speed trajectories of the three micro-vehicles. The leader travels at a speed of $1m/s$ during the whole process. Vehicle2 and vehicle3 have a significant acceleratation when joining tail and then track the speed of the preceeding vehicle.

Moreover, the experiment verifies that the proposed maneuver management framework could run on the VCU of the micro-vehicles, so it satisfies the computational requirements of real-time.

\section{CONCLUSIONS}

This paper is mainly motivated by the challenges linked to the scalability of maneuvers of vehicle platoon in the practical applications. 

In this paper, an extendable two-dimensional maneuver management framework with fault-tolerant mechanism is introduced on the basis of the hierarchical architecture for the platoon control system. The one advantage of this framework is extendable. The maneuvers under this framework could be extended without revising the existing strategies since each maneuver is independent of each other. The other advantage of this framework is universal. The management framework equipped in every participating vehicle is exactly the same, so the vehicle with the proposed framework could switch to any role of the platoon as the management strategy commands. Additionally, the fault-tolerant mechanism is added to the framework as a maneuver for handling the hardware failures. Functionality degradation maintain the safefy until the driver takes over.

A simulation platform for a five-vehicle platoon driving on the highway is established. The result of the overall platoon simulation of integrated maneuvers show that the proposed two-dimensional framework could effectively deal with these maneuvers effectively and the framework is applicable to each role in the platoon. And the comparison of the results with and without the fault-tolerant mechanism when the hardware failures verifies the key role of the functionality degradation for maintaining the safety of the platoon. Moreover, a real-world experiment using three micro-vehicles is implemented in Tsinghua University. The results further validate the effectiveness of the management strategies and the proposed framework could satisfy the computational real-time requirements. Future work will seek to verify the effect of the maneuver management framework and management strategies for other maneuvers.

\addtolength{\textheight}{-12cm}   % This command serves to balance the column lengths

%%%%%%%%%%%%%%%%%%%%%%%%%%%%%%%%%%%%%%%%%%%%%%%%%%%%%%%%%%%%%%%%%%%%%%%%%%%%%%%%

%%%%%%%%%%%%%%%%%%%%%%%%%%%%%%%%%%%%%%%%%%%%%%%%%%%%%%%%%%%%%%%%%%%%%%%%%%%%%%%%

%%%%%%%%%%%%%%%%%%%%%%%%%%%%%%%%%%%%%%%%%%%%%%%%%%%%%%%%%%%%%%%%%%%%%%%%%%%%%%%%
% \section*{APTENDIX}


%%%%%%%%%%%%%%%%%%%%%%%%%%%%%%%%%%%%%%%%%%%%%%%%%%%%%%%%%%%%%%%%%%%%%%%%%%%%%%%%

\bibliographystyle{vancouver}

\bibliography{myrefs}
\end{document}